\documentclass[journal]{IEEEtran}
\usepackage{graphicx}
\usepackage{amsmath}
\usepackage{amsfonts}
\usepackage{breqn}
\usepackage{multirow}
\usepackage{tikz}
\usetikzlibrary{patterns}
\usetikzlibrary{arrows}
\usetikzlibrary{shapes,snakes}
\usetikzlibrary{backgrounds,fit,decorations.pathreplacing}
\usepackage{graphicx}
\usepackage{caption}
\usepackage{subcaption}
\usepackage{pgfplots}
\pgfplotsset{grid style={dashed, gray}}
\usetikzlibrary{shapes,arrows,patterns}
\graphicspath{ {./images/} }
\usepackage{authblk}

\ifCLASSINFOpdf
\else
\fi

\begin{document}
\bstctlcite{IEEEexample:BSTcontrol}
%
\title{Energy Efficient Routing and Network Coding  in Core Networks}
%
%
%

\author[]{Mohamed~Musa}

\author[]{Taisir~Elgorashi}
\author[]{Jaafar~Elmirghani}
\affil[]{School of Electronic and Electrical Engineering, University of Leeds, LS2 9JT, UK}



\maketitle

\begin{abstract}
We propose network coding as an energy efficient data transmission technique in core networks with non-bypass and bypass routing approaches. The improvement in energy efficiency is achieved through reduction in the traffic flows passing through intermediate nodes. A mixed integer linear program (MILP) is developed to optimize the use of network resources, and the results show that our proposed network coding approach introduces up to 33\% power savings for the non-bypass case compared with the conventional architectures. For the bypass case, 28\% power savings are obtained considering futuristic network components’ power consumption. A heuristic based on the minimum hop count routing shows power savings comparable to the MILP results. Furthermore, we study how the change in network topology affects the savings produced by network coding. The results show that the savings are proportional to the average hop count of the network topology. We also derive power consumption analytic bounds and closed form expressions for networks that implement network coding and thus also verify the results obtained by the MILP model.

\end{abstract}


\IEEEpeerreviewmaketitle

\section{Introduction}

%
%
%
%
\IEEEPARstart{T}{he} exponential growth in the use of ICT has intensified the interest in energy efficiency approaches. The current ICT systems are not energy efficient, and the inefficiency gap is continuously  expanding. This inherent engineering problem is steering attention towards new designs with energy efficiency objectives in mind. Economic factors, especially after the increase in energy prices, have increased the pressure to push the efficiency to its limits. These factors alongside the ecological factors will probably be one of the important bottlenecks to ICT expansion. The total emissions of ICT are estimated in The GeSI SMARTer 2030 report \cite{GeSI2015} to constitute 1.97\% of all emissions by the year 2030.\\
Greening the Internet was first introduced in 2003 \cite{Gupta2003} as a concept. The authors of \cite{Zhang2010} provide a survey of energy efficiency improvement approaches used in optical networks. Previously, our work has addressed the core network energy efficiency   with renewable energy sources \cite{dong2011ip}, core networks with communication between data centers \cite{Dong2011} and  data centres with dis-aggregated servers \cite{ali2017future}, design and optimization of physical topology \cite{Dong2012}, the efficient distribution of clouds \cite{LaweyCloud}, future energy efficient high definition TV \cite{Niema2014}, Peer to Peer content distribution \cite{Lawey2014, al2019energy}, big data management \cite{hadi2019patient, hadi2018big,al2017energy}, energy efficient embedding of virtual network \cite{Nonde2015}, network function virtualization \cite{al2019optimized} and the improvement in energy efficiency resulting from multiple combined methods in the GreenTouch GreenMeter \cite{elmirghani2018greentouch, musa2018bounds2}. \\
In 2000, Ahlswede et. al proposed the idea of network coding in wireless networks for multicast connections \cite{ahlswede2000},  achieving a throughput improvement by using all network nodes, which participate in the coding process as opposed to the conventional routing approach. This throughput improvement means a possible opportunity for energy efficiency improvement. In \cite{fragouli2007b} a good review of the applications of network coding is provided. However, most of the work on network coding focused on wireless networks. The most notable work in optical networks is in  passive optical networks \cite{Belzner2009, Fouli2011, Liu2012} or protection for mesh optical networks \cite{Aly2008, Kamal2011, babarczi2012optimal,chen2017linear} and our previous work \cite{musa2018bounds, musa2017energy}. These applications of network coding rely on the multicast advantage of the PON and the multipath requirement of the protection mechanism. The rest of the efforts addressed the impact of using multicast enabled optical  switches for network coding \cite{Manley2008, Manley2010} as well as  the requirement to do network coding in the physical layer \cite{LiuZ2012}. The authors in \cite{kim2007genetic} and \cite{hu2014evolutionary} proposed approaches to reduce the network coding resource in the network, and in \cite{cui2015optimization} the authors proposed approaches that allow fairly general coding across flows. \\
In this paper we expand on our previous work in \cite{MusaICTON, musa2016energy} and present a detailed investigation of a new energy efficient routing and network coding mechanism in  unicast settings where opposite flows passing through intermediate nodes get encoded using a simple xor operation before being sent back to their destination. This calls for simple hardware changes at the IP layer only. \\
Although the idea we present is similar in principle to the work of \cite{Katti2006} in wireless networks, our work is different because a wired implementation of network coding is very different from its wireless counterpart. This is because the benefit in the wireless case is due to the shared broadcast domain between nodes, whereas the benefit in the wired case is less clear since wired switched networks have a smaller collision domain between neighbours and support full duplex communication, and therefore need to be explored and understood. We use a MILP model followed by a heuristic to determine the performance of our new network coding approach, and similar to our approach in \cite{musa2018bounds, musa2017energy}, we provide analytic bounds and closed form expressions to validate the MILP model and the heuristic. We consider different approaches for implementing network coding to address the possibility of packet size mismatch, and also our work makes a sensible quantified  estimate of the benefit that could be achieved in practice. \\
Section II of this paper introduces network coding and proposes a novel node architecture to support network coding in non-bypass IP over WDM networks. In Section III the new approach is modelled using mixed integer linear programming (MILP), with the analysis of results presented in Section IV. In Section V we compare two approaches of network coding with partitioning and with zero padding of flows, and  we investigate implementing network coding in bypass IP over WDM networks in Section VI.  Section VII presents power consumption analytic bounds and closed form expressions of the conventional and the network coding scenarios. Finally, Section VIII concludes the paper. 
\section{Concept description}
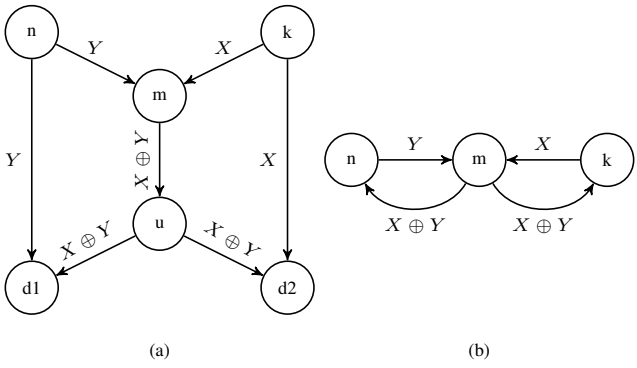
\begin{figure}[]
 \centering
	\begin{tikzpicture}[>=stealth',semithick,auto, font=\scriptsize, scale=0.85]
    \tikzstyle{obj}  = [circle, minimum width=20pt, draw, inner sep=1pt]
    \tikzstyle{every label}=[font=\bfseries]      
        		\node[obj] (n) at (5,-4) {n};
		\node[obj] (m) at (7,-4) {m};
    		\node[obj] (k) at (9,-4) {k};
   	 \path[->]   (n)    edge                node[swap,above]  {$Y$}       (m);
  	  \path[->]   (k)    edge                node[swap,above]  {$X$}       (m);
	 \path[->]   (m)    edge [bend right=60]               node[swap,below]  {$X \oplus Y$}       (k);
	 \path[->]   (m)    edge [bend left=60]               node[swap,below]  {$X \oplus Y$}       (n);  
  	\node[obj] (n1) at (0,-2) {n};
	\node[obj] (m1) at (2,-3) {m};
    	\node[obj] (k1) at (4,-2) {k};
	 \node[obj] (n2) at (0,-6) {d1};
	\node[obj] (m2) at (2,-5) {u};
    	\node[obj] (k2) at (4,-6) {d2};
	
	\path[->]   (n1)    edge                node[swap,above]  {$Y$}       (m1);
	\path[->]   (k1)    edge                node[swap,above]  {$X$}       (m1);
	\path[->]   (m1)    edge                node[ above, rotate=90]  {$X \oplus Y$}       (m2);
	\path[->]   (m2)    edge                node[swap, above, rotate=30]  {$X \oplus Y$}       (n2);
	\path[->]   (m2)    edge                node[swap, above, rotate=-30]  {$X \oplus Y$}       (k2);
	\path[->]   (n1)    edge                node[swap]  {$Y$}       (n2);
	\path[->]   (k1)    edge                node[swap]  {$X$}       (k2);
	
		\node[font=\scriptsize] (a) at (2,-7) {(a)};
		\node[font=\scriptsize] (b) at (7,-7) {(b)};
	\end{tikzpicture}
\caption{The butterfly network (a) and the three nodes network (b)}
\label{fig:Three_nodes}
\end{figure}

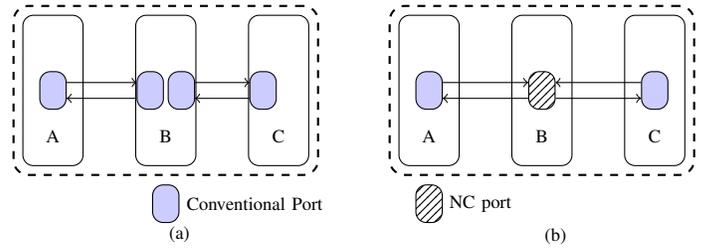
\begin{figure}
\centering
\pgfdeclarelayer{background}
\pgfdeclarelayer{foreground}
\pgfsetlayers{background,main,foreground}
  \begin{tikzpicture}[node distance=1.5 cm]
  \tikzstyle{surround} = [thick,draw=black,rounded corners=2mm, dashed]
  \tikzstyle{BigNode}=[draw, thin,minimum width=0.8cm,minimum height=2cm, rounded corners]
  \tikzstyle{SmallBlock} = [draw,fill=blue!20, thin,minimum width=0.35cm,minimum height=0.5cm, rounded corners]
\node [BigNode](a){};
\node [BigNode, right of=a](b){};
\node [BigNode, right of =b](c){};
\path (a.south) + (0,+0.4)node (labela) {\scriptsize A};
\path (b.south) + (0,+0.4)node (labelb) {\scriptsize B};
\path (c.south) + (0,+0.4)node (labelc) {\scriptsize C};
\node [BigNode](a2) at (5,0){};
\node [BigNode, right of=a2](b2){};
\node [BigNode, right of =b2](c2){};
\path (a2.south) + (0,+0.4)node (labela2) {\scriptsize A};
\path (b2.south) + (0,+0.4)node (labelb2) {\scriptsize B};
\path (c2.south) + (0,+0.4)node (labelc2) {\scriptsize C};
\node [SmallBlock](x){};
\path (b.180)+(+0.2,0) node (y) [SmallBlock] {};
\path (b.0)+(-0.2,0) node (y2) [SmallBlock] {};
\node [SmallBlock, right of =y](z){};
\draw [->] (x.30) -- (y.150);
\draw [->] (y.210) -- (x.330);
\draw [->] (y2.30) -- (z.150);
\draw [->] (z.210) -- (y2.330);
\node [SmallBlock](x1) at (5,0){};
\node [SmallBlock, right of=x1, pattern=north east lines,pattern color=black](y1){};
\node [SmallBlock, right of =y1](z1){};
\draw [->] (x1.30) -- (y1.150);
\draw [->] (z1.150) -- (y1.30);
\draw [->] (y1.210) -- (x1.330);
\draw [->] (y1.330) -- (z1.210);
\path (b.south)+(+0,-0.5) node (leg) [SmallBlock] {};
\path (leg.east) +(+1,0) node (leg) [] {\scriptsize Conventional Port};
\path (a2.south)+(+0,-0.5) node (leg2) [SmallBlock, pattern=north east lines,pattern color=black] {};
\path (leg2.east)+(+0.5,0) node (leg2)  {\scriptsize NC port};
\path (leg2.south)+(+1,-0.2) node (leg2)  {\scriptsize (b)};
\path (leg.south)+(-1,-0.2) node (leg2)  {\scriptsize (a)};
\begin{pgfonlayer}{background} 
    \node[surround] (background) [fit = (a) (c) (b)] {};
    \node[surround] (background) [fit = (a2) (b2) (c2)] {};
    
    \end{pgfonlayer}
\end{tikzpicture}
\caption{Conventional architecture (a), and Network coding architecture (b)}
\label{fig:Classical_NC}
\end{figure}

\begin{figure}
\centering
\pgfdeclarelayer{background}
\pgfdeclarelayer{foreground}
\pgfsetlayers{background,main,foreground}
\begin{tikzpicture}[node distance=3 cm]
\tikzstyle{surround} = [thick,draw=black,rounded corners=2mm, dashed]
\tikzstyle{BigNode}=[draw, thick,minimum width=2cm,minimum height=3cm, dashed]
\tikzstyle{SmallBlock} = [draw, thin,minimum width=0.25cm,minimum height=0.1cm]
\tikzstyle{Storage} = [draw, thin,minimum width=0.7cm,minimum height=0.3cm]

\node [BigNode](a){};
\node [BigNode, right of=a](b){};
\node [BigNode, right of =b](c){};
\path (a.45) + (-.275,0.3) node[SmallBlock] (tx1) {\scriptsize Tx};
\path (a.45) + (-.275,-0.15) node[SmallBlock] (rx1) {\scriptsize Rx};
\path (c.135) + (+.25,0.3) node[SmallBlock] (tx2) {\scriptsize Tx};
\path (c.135) + (+.25,-0.15) node[SmallBlock] (rx2) {\scriptsize Rx};
\path (b.135) + (+.25,0.3) node[SmallBlock] (rx3) {\scriptsize Rx};
\path (b.45) + (-.25,0.3) node[SmallBlock] (rx4) {\scriptsize Rx};
\path (b) + (0,-0.7) node[SmallBlock] (tx3) {\scriptsize Tx};
\path (a)+(0,0.3)  node[Storage] (Str1) {\scriptsize Storage};
\path (b) +(0,0.4) node[Storage] (Str2) {\scriptsize Storage};
\path (c) +(0,0.3) node[Storage] (Str3) {\scriptsize Storage};

\path (a) +(0,-0.5) node[Storage] (xor1) {\scriptsize xor};
\path (b)  +(0,-0.2) node[Storage] (xor2) {\scriptsize xor};
\path (c)  +(0,-0.5) node[Storage] (xor3) {\scriptsize xor};
\path (b)  +(0,-1.2) node[Storage] (splitter) {\scriptsize splitter};

\path (a.south) + (0,-0.1) node (labela) {\scriptsize node (n)};
\path (b.south) + (0,-0.1) node (labelb) {\scriptsize intermediate node (m)};
\path (c.south) + (0,-0.1) node (labelc) {\scriptsize node(k)};

\draw [->] (rx1.-90)  |-  (xor1);
\draw [->] (Str1.-90) -- (xor1);

\draw [->] (rx3) |- (Str2.180);
\draw [->] (rx4)|- (Str2.0);
\draw [->] (Str2)-- (xor2);
\draw [->] (xor2) -- (tx3);

\draw [->] (rx2.-90)  |-  (xor3);
\draw [->] (Str3.-90) -- (xor3);

\draw [->] (tx2)  -- node[above] {\scriptsize $x$} (rx4);
\draw [->] (tx1)  --node[above] {\scriptsize $y$}  (rx3);
\draw [->] (tx3)  -- (splitter);

\draw (splitter) edge[out=180,in=0,->] node[midway,below, rotate=-60] {\scriptsize $x+y$} (rx1) ;
\draw (splitter) edge[out=0,in=180,->] node[midway,rotate=60, below] {\scriptsize $x+y$} (rx2);
\path (a.180) +(-0.4,-0.5) node[] (x) {\scriptsize x};
\path (c.0) +(0.4,-0.5) node[] (y) {\scriptsize y};
\draw [->] (xor1.180)  --  (x.0);
\draw [->] (xor3.0)  --  (y.180);
\end{tikzpicture}
\caption{Network coding node architecture}
\label{fig:NodeArch}
\end{figure}
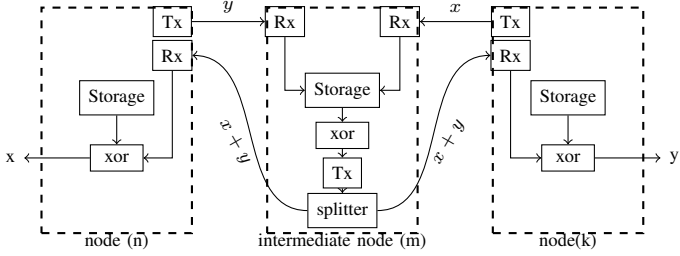

Network coding is commonly illustrated by the well-known butterfly network shown in Fig. \ref{fig:Three_nodes} (a). Here the sources $k$ and $n$ want to multicast two units of information, $X$ and $Y$, to two receivers, $d2$ and $d1$, respectively. Assuming all links have a unit capacity in the direction shown in the Figure, tracing the flows will lead to a bottleneck in link $(m,u)$. Instead of doubling the link capacity to resolve this bottleneck, the link capacity can be shared by the two flows by encoding the flows by an XOR operation in node $m$ and multicasting the encoded flow to $d1$ and $d2$ from node $u$.
Fig. \ref{fig:Three_nodes} (b) shows a special case of the butterfly network where each of the nodes pairs $(n,d1)$, $(m,u)$ and $(k,d2)$ are considered as a single node, where the links connecting them are regarded as storage rather than communication links, this converts the butterfly network to a 3 node network $n,m$ and $k$. This special butterfly can perform network coding in a unicast scenario. Suppose node $k$ wants to transmit the information units $X$ to node $n$, and node $n$ wants to transmit $Y$ to node $k$. Both flows will go through the intermediate node $m$, which can combine the flows using the XOR gate and multicast back the encoded flow to both end nodes. Each node retrieves the information unit intended to it by XOR coding the received unit with the stored information unit. In the middle node where coding is performed, we encounter resource savings.\\
The new approach is described in Fig. \ref{fig:Classical_NC}. At end nodes where flows originate or terminate, the same ports as the conventional routing approach will be used. At intermediate nodes, however, two conventional ports are replaced by a single port that implements the coding functionality. In this paper, the term conventional port (architecture) is used to mean the current implementation, and the NC port (architecture) for the network coding enabled IP port. The coding functionality used here is the mere modulo-2 addition (XOR). For the three nodes example shown in Fig. \ref{fig:Classical_NC} ,four conventional ports are reduced to a single NC port and two conventional ports. The amount of power savings depends on the ratio of the power consumed by the two types of ports and the assignment of flows to their routes.
The IP over WDM network is the current core networks implementation; traffic collected from edge routers is passed to core IP routers, and the optical layer provides transport over the high bandwidth optical fibres. Transponders convert the electric signal into optical and optical cross connects provide the optical layer switching. In each fiber,  multiple wavelengths are used through wavelength division multiplexing with the aid of  multiplexers/demultiplexers. Erbium doped fiber amplifiers (EDFAs) provide optical signal amplification to support transmission over long distances. The existence of the IP and the WDM layers presents two approaches to route the traffic. The first is referred to as the non-bypass approach which routes traffic from its originating IP router to its destination IP router passing through each IP router in intermediate nodes. The second approach bypasses intermediate IP routers and is hence referred to as the bypass approach. \\
Our proposed architecture of the network coding enabled IP over WDM node is illustrated in Fig. \ref{fig:NodeArch}. The new NC enabled port uses two receivers to receive the two flows that will be encoded and transmitted using the single transmitter. A coupler is used to multicast the flow from the single transmitter to the two neighbouring nodes.  Storage is needed to perform the synchronization task prior to encoding. Note that, even if we perform the synchronization in the IP layer, different data flows may not arrive at the intermediate node at the same time. Therefore, storage is essential here. The XOR unit is used to encode the flows. An amplifier is used for splitting power loss compensation and to support long distance optical transmission. Note that an alternative architecture that uses an optical implementation of the XOR gate is a straightforward extension of Fig. \ref{fig:NodeArch}. The storage unit in source nodes and destination nodes are used to store the original data and the XOR is used to decode the received encoded flow. The cost of the additional storage is mitigated by performing the coding in the IP layer and therefore storage and the xor functionality can be included as part of the available processing power. 

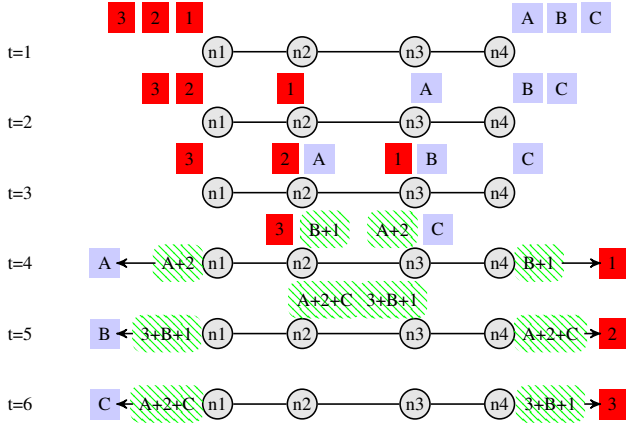
\begin{figure}
\centering
\begin{tikzpicture}[>=stealth',semithick,auto, scale=0.75]
	\tikzstyle{rectRed}=[fill=red,rectangle,minimum width=3mm,minimum height=3mm,  font=\scriptsize]
	\tikzstyle{rectBlue}=[fill=blue!20,rectangle,minimum width=3mm,minimum height=3mm,  font=\scriptsize]
    \tikzstyle{subj} = [circle, minimum width=10pt, fill, inner sep=1pt]
    \tikzstyle{obj}  = [circle, minimum width=10pt, draw, inner sep=1pt,  font=\scriptsize]
    \tikzstyle{dc}   = [circle, minimum width=10pt, draw, inner sep=1pt, path picture={\draw (path picture bounding box.south east) -- (path picture bounding box.north west) (path picture bounding box.south west) -- (path picture bounding box.north east);}]
	\tikzstyle{coded}=[fill=green,rectangle,pattern=north west lines, pattern color=green,minimum width=5mm,minimum height=5mm, font=\scriptsize, rounded corners]
	\tikzstyle{every label}=[font=\bfseries]
    
\node[] (t1) at (-3,-1.25){\scriptsize t=1};
\node[] (t2) at (-3,-2.5){\scriptsize t=2};
\node[] (t3) at (-3,-3.75){\scriptsize t=3};
\node[] (t4) at (-3,-5){\scriptsize t=4};
\node[] (t5) at (-3,-6.25){\scriptsize t=5};
\node[] (t6) at (-3,-7.5){\scriptsize t=6};

    \foreach \x in {1,...,6}{
    
    	\node[obj, fill=black!10] (a) at (0.5,0-1.25*\x) {n1};
	\node[obj,fill=black!10] (b) at (2,0-1.25*\x) {n2};
    	\node[obj,fill=black!10] (c) at (4,0-1.25*\x) {n3};
     	\node[obj,fill=black!10] (d) at (5.5,0-1.25*\x) {n4};
     
    \path[-]   (a)    edge                node[swap]  {}       (b);
    \path[-]   (b)    edge                node[swap]  {}       (c);
    \path[-]   (c)    edge                node[swap]  {}       (d);
    }
    \node at (0,-0.65) [rectRed] {1};
    \node at (0,-1.9) [rectRed]{2};
    	\node at (-0.6,-1.9)[rectRed] {3};
    \node at (0,-3.15) [rectRed] {3};
    \node at (-0.6,-0.65) [rectRed] {2};
     \node at (-1.2,-0.65) [rectRed] {3};
     
    \node at (6,-0.65) [rectBlue] {A};
    \node at (6.6,-0.65) [rectBlue] {B};
    \node at (7.2,-0.65) [rectBlue] {C};
    \node at (1.8,-1.9) [rectRed]{1};
    \node at (4.2,-1.9) [rectBlue] {A};
    \node at (6,-1.9)[rectBlue] {B};
    \node at (6.6,-1.9) [rectBlue]{C};
    \node at (6,-3.15) [rectBlue]{C};
    
    \node at (1.7,-3.15) [rectRed] {2};
    \node at (2.3,-3.15) [rectBlue] {A};
    \node at (3.7,-3.15) [rectRed] {1};
    \node at (4.3,-3.15) [rectBlue]{B};
         
    \node at (1.6,-4.4) [rectRed]{3};
    \node at (2.4,-4.4) [coded] {B+1};
    \node at (3.6,-4.4) [coded] {A+2};
    \node at (4.4,-4.4) [rectBlue] {C};  
         
              \node at (-0.2,-5) [coded]  {A+2}; 
                           \node at (-1.5,-5) [rectBlue] {A};  
                           \path [->] (-0.6,-5) edge (-1.3,-5) ;
                           
                    \node at (6.2,-5) [coded]  {B+1}; 
                           \node at (7.5,-5) [rectRed] {1};  
                           \path [->] (6.6,-5) edge (7.3,-5) ;
         
     \node at (2.4,-5.65) [coded]  {A+2+C};
     \node at (3.6,-5.65)[coded]  {3+B+1};
     
                  \node at (-0.4,-6.25) [coded]  {3+B+1}; 
                           \node at (-1.5,-6.25)[rectBlue]{B};  
                           \path [->] (-1,-6.25) edge (-1.3,-6.25) ;
                    \node at (6.4,-6.25) [coded] {A+2+C}; 
                           \node at (7.5,-6.25) [rectRed] {2};  
                           \path [->] (7,-6.25) edge (7.3,-6.25) ;
                           
   	\node at (-0.4,-7.5) [coded] {A+2+C}; 
                           \node at (-1.5,-7.5) [rectBlue]{C};  
                           \path [->] (-1,-7.5) edge (-1.3,-7.5) ;
                    \node at (6.4,-7.5) [coded]  {3+B+1}; 
                           \node at (7.5,-7.5) [rectRed] {3}; 
                     \path [->] (7,-7.5) edge (7.3,-7.5) ;

\end{tikzpicture}
\caption{Network coding at two intermediate nodes}
\label{fig:NC_Two_intermediate}
\end{figure}

Network coding can be performed at all intermediate nodes of bidirectional flows. Fig. \ref{fig:NC_Two_intermediate} illustrates how the network coding is performed in a bidirectional flow traversing two intermediate hops. We consider an example where node $n1$ needs to send packets $1$, $2$ and $3$ to node $n4$ and node $n4$ needs to send packets $A$, $B$, and $C$ to node $n1$. The packets being sent, received or processed by a node are shown above the node.
\begin{itemize}
    \item At $t=1$, packets ($1$) and ($A$) are being prepared for transmission by ($n1$) and ($n4$), respectively.
    \item At $t=2$, packets ($1$) and ($A$) arrive at their neighbouring intermediate nodes, meanwhile packets ($2$) and ($B$) are being prepared by nodes ($n1$) and ($n4$), respectively. As intermediate nodes ($n2$) and ($n3$) receive only a single packet from one side no coding takes place and the packets are forwarded to the next node.
    \item At $t=3$, packets ($1$) and ($B$) arrive at node ($n3$) and packets ($A$) and ($2$) arrive at node ($n2$). Nodes ($n2$) and ($n3$) encode their packets and produce ($2+A$) and ($1+B$) respectively. The encoded packets are multicast to the two neighbouring nodes (e.g. for node ($n2$) it goes to nodes ($n1$) and ($n3$)). 
    \item At $t=4$, end node ($n1$) receives packet ($A+2$) and decodes it using the stored packet ($2$). The same applies to node ($n4$) decoding packet ($B+1$) with the stored packet ($B$). Meanwhile, intermediate nodes ($n2$) and ($n3$) receive packets ($B+1$) and ($A+2$). First they decode them using the stored packets ($1$) and ($A$), to produce packets ($B$) and ($2$), ($n2$) then encodes ($B$) with ($3$) to produce ($B+3$), and ($n3$) encodes ($C$) with ($2$) to produce ($C+2$), which both are multicast to their two intended neighbours. 
    \item At the remaining time slots, end nodes decode their received coded packets using their stored packets in a straightforward manner. 
\end{itemize}

 We assume that packets are fully synchronized at reception (buffering can be used to help the synchronisation process). Nodes decide which packets are encoded in a given packet from a metadata, which is carried in the header and negligible in size compared to packet sizes. Another possibility is to do coding on established paths in a circuit switched network, where paths are decided, and hence encoding possibilities, will be decided beforehand. We also assume the control and management overhead is negligible since it can be performed as a result of a software upgrade to the system. 
 
\section{The MILP model} \label{milp}
This section describes the  MILP model developed whose objective is to minimize the total power consumption of a non-bypass IP over WDM network with the aforementioned network coding implementation by optimizing the routes each demand takes, and optimizing the number and location of conventional and NC ports, for a given network topology and demands matrix. Below are lists of sets, parameters and variables defined in the MILP model:

\begin{center}
\begin{table}[h]
\caption{The sets of the MILP model and their description}
 \small
 \def\arraystretch{1.5}\tabcolsep=10pt
    \centering
    \begin{tabular}{c  p{6 cm} } 
    \hline
    Set & description \\
 \hline \hline
$\mathcal{N}$ & Set of network nodes  \\ 
 \hline
 $\mathcal{N}_m$ & Set of neighbouring nodes connected to node $m$  \\ 
 \hline
\end{tabular}
    \label{NCvariables}
\end{table}
\end{center}

\begin{center}
\begin{table}[h]
\caption{Model parameters and their description}
 \small
 \def\arraystretch{1.5}\tabcolsep=10pt
    \centering
    \begin{tabular}{c  p{6 cm} } 
    \hline
    Parameter & Parameter's description \\
 \hline \hline
$\lambda^{sd}$ & The traffic volume of the demand $(s,d)$ in multiples of wavelengths  \\ 
 \hline
$L^{mn}$ & The length of the physical link  $(m,n)$ \\ 
 \hline
$B$ & The wavelength capacity in Gbps\\ 
 \hline
$W$ & wavelengths count in each fibre\\ 
 \hline
${s,d}$ & A tuple denoting the demand source and destination.\\ 
 \hline
${m,n}$ & Denotes end nodes of a given physical link\\ 
 \hline
$p_p$ &  Conventional router port power consumption\\ 
 \hline
$p_x$ & Network coding based router port power consumption. \\ 
 \hline
 $p_t$ & Transponder power consumption. \\ 
 \hline
 $p_e$ & EDFA power consumption. \\ 
 \hline
 $p_o$ & Optical switch power consumption. \\ 
 \hline
 $p_{md}$ & Multiplexer/demultiplexer power consumption. \\ 
 \hline
 $S$ & Distance required between two neighbouring EDFAs\\ 
 \hline
\end{tabular}
    \label{NCparameters}
\end{table}
\end{center}

\begin{center}
\begin{table}[h]
\caption{Model variables and their description}
 \small
 \def\arraystretch{1.5}\tabcolsep=10pt
    \centering
    \begin{tabular}{c  p{6 cm} } 
    \hline
    Variable & Variable description \\
 \hline \hline
$P_T$ & The total power consumption of the whole network (W)\\ 
 \hline
$w^{sd}_{m,n}$ & The traffic flow between source and destination node pair $(s,d)$ that traverses the optical link $(m,n)$, in Gbps\\ 
 \hline
$b^{sd}_{m,n}$ & Binary equivalent of $w^{sd}_{m,n}$, $b^{sd}_{m,n}=1$ if $w^{sd}_{m,n}> 0$, $b^{sd}_{m,n}=0$ otherwise\\ 
 \hline
$c^{sd}_{nmk}$ & $c^{sd}_{nmk}=1$ if demand $(s,d)$ traverses links $(n,m)$ and $(m,k)$ (i.e. bidirectional traffic flows can be encoded in node $m$) \\ 
 \hline
$Y_m$ & Number of conventional ports at node $m$\\ 
 \hline
$X_m$ & Number of NC ports at node $m$\\ 
 \hline
$X^m_{nk}$ & Number of NC ports at node $m$ catering for the traffic encoding between the node pair $(n,k)$\\ 
 \hline

$A_{mn}$ & The total number of EDFAs on a physical link $(m,n)$. This is calculated using $A_{mn}=\lfloor L_{mn}/S-1\rfloor$.  $S$ is the separating distance between neighbouring amplifiers. \\ 
 \hline
$f_{mn}$ & The total number of fibres used on physical link $(m,n)$\\ 
 \hline
$w_{mn}$ & The total traffic flow carried on physical link $(m,n)$\\ 
 \hline
$w^m_{nk}$ & The total traffic flow between node pair $(n,k)$ that goes through the intermediate node $m$\\ 
 \hline
$Npo_{mn}$ & The number of conventional router ports where the traffic starts at source node $m$ and terminates in a node $n$\\ 
 \hline
$Npi_{mn}$ & The number of conventional router ports where the traffic terminates at destination node $n$ and starts in a node $m$\\ 
 \hline
 $Np_{mn}$ & The number of conventional router ports at node $m$ interfacing node $n$\\ 
 \hline
\end{tabular}
    \label{NCvariables}
\end{table}
\end{center}
The model is defined as follows:\\
\textbf{Objective}: minimize the power consumption of the network coding enabled IP over WDM network:
\begin{align*} \label{obj_func}
P_T=\sum _{ m\in \mathcal{N} }\bigg( p_p Y_{ m }+p_x X_{ m }+p_o +p_{md}\\
+ \sum _{ n \in \mathcal{N}_m }{\left( p_t w_{ mn }+p_e A_{ mn } f_{ mn } \right) }\bigg). 
\end{align*}
The power consumption outlined in the objective function is made of the following components:
\begin{itemize}
    \item  The power consumption of conventional router ports
    \begin{equation}
    \sum _{ m\in \mathcal{N} } p_p Y_{m}.
    \end{equation}
    \item The power consumption of NC router ports
    \begin{equation}
    \sum _{ m\in \mathcal{N} } p_x X_{m}.
    \end{equation}
    \item  The power consumption of optical switches
    \begin{equation}
    \sum _{ m\in \mathcal{N} }p_o.
    \end{equation}
    \item  The total power consumption contribution from multiplexers and demultiplexers
    \begin{equation}
    \sum _{ m\in \mathcal{N} }p_{md}.
    \end{equation}
    \item  The total power consumption contribution from transponders 
    \begin{equation}
    \sum _{ m\in \mathcal{N} }\sum _{ n \in \mathcal{N}_m } p_t w_{ mn }.
    \end{equation}
    \item  The power consumption of all the EDFAs in the network
    \begin{equation}
    \sum _{ m\in \mathcal{N} }\sum _{ n \in \mathcal{N}_m } p_e A_{ mn } f_{ mn }.
    \end{equation}
\end{itemize}
Subject to: 
\begin{equation} \label{flow_conservation}
\sum _{ n\in \mathcal{N}_m } b_{ mn }^{ sd }-\sum _{ n\in \mathcal{N}_m }b_{ nm }^{ sd } =\left\{ \begin{array}{lr} 1 & :m=s \\ -1 & :m=d \\ 0& :otherwise \end{array} \right\}
\end{equation}
\centerline{$\forall s,d,m \in \mathcal{N}$}.
 Constraint (\ref{flow_conservation}) represents the flow conservation constraint. It ensures traffic routing from source to destination. 
 
\begin{equation} \label{calc_flow}
 w_{ mn }^{ sd } =\lambda ^{sd} b_{ mn }^{ sd },
\end{equation}
\centerline{$\forall s,d \in \mathcal{N}, \forall m \in \mathcal{N}, n \in \mathcal{N}_m$}.
Constraint (\ref{calc_flow}) calculates the flow of a traffic demand that traverses a link based on the binary variable  $b_{mn}^{sd}$. 
\begin{equation} \label{calc_flow2}
 w_{ mn } =\sum_{s \in \mathcal{N}}\sum_{d \in \mathcal{N}:s \ne d}w_{ mn }^{ sd },
\end{equation}
\centerline{$\forall m \in \mathcal{N},n \in \mathcal{N}_m$}.
 Constraint (\ref{calc_flow2}) calculates the aggregate traffic on a given link which is represented by the total flow of all demands passing through that link. 

\begin{equation} \label{capacity_cons}
\sum_{ s\in \mathcal{N} } \sum_{\substack{ d \in \mathcal{N} \\ s \neq d }} w_{ mn }^{ sd } \le W B f_{ mn },
\end{equation}
\centerline{$\forall s,d,m \in \mathcal{N},  n \in \mathcal{N}_{ m }$}.
The link capacity conservation constraint is ensured by constraint (\ref{capacity_cons}), where the total flow on a link must not surpass the total capacity of all fibres on that  link.

\begin{equation} \label{NumOutPorts}
Npo_{mn}= \sum _{ s\in \mathcal{N} } \sum _{\substack{ d\in \mathcal{N}\\ s\neq d \\s=m}}  \frac{w_{mn}^{ sd}}{B}.
\end{equation}
\begin{equation} \label{NumInPorts}
Npi_{mn}= \sum _{ s\in \mathcal{N} } \sum _{\substack{ d\in \mathcal{N}\\ s\neq d \\d=m}}  \frac{w_{mn}^{ sd}}{B}.
\end{equation}
Equation (\ref{NumOutPorts}), and (\ref{NumInPorts}) are used in the MILP implementation instead of the more accurate $\lceil \sum _{ s\in \mathcal{N} } \sum _{\substack{ d\in \mathcal{N}\\ s\neq d \\s=m}}  \frac{w_{mn}^{ sd}}{B} \rceil$ and  $\lceil \sum _{ s\in \mathcal{N} } \sum _{\substack{ d\in \mathcal{N}\\ s\neq d \\d=m}}  \frac{w_{mn}^{ sd}}{B} \rceil$ respectively which select the next integer greater than a given real value. This is 
due to relaxation of the MILP model.\\
Now that the number of conventinoal ports leaving and entering a node are determined by equations (\ref{NumOutPorts}) and (\ref{NumInPorts}) respectively, and noting that router ports have a pair (i.e. Tx and Rx components), then under the asymmetric traffic the larger of the outgoing and the incoming traffic at a node determines the number of conventional router ports needed as shown in Equation \ref{NumPorts1}. 
\begin{equation} \label{NumPorts1}
Np_{mn} =\max(Npo_{mn},Npi_{mn})
\end{equation}
\centerline{$\forall m \in \mathcal{N}, n \in \mathcal{N}_m$}
The total number of conventional ports is then given by:
\begin{equation} \label{NumPorts}
Y_m=\sum_{n \in N_m}Np_{mn},
\end{equation}
\centerline{$\forall m \in \mathcal{N}$}.
\vspace{5mm}
\begin{equation} \label{CwCalc1}
c_{nmk }^{ sd } \le b_{ nm }^{ sd },
\end{equation}
\begin{equation} \label{CwCalc2}
c_{nmk }^{ sd } \le b_{ mk }^{ sd },
\end{equation}
\begin{equation} \label{CwCalc3}
c_{nmk }^{ sd } \ge b_{ nm }^{ sd } + b_{ mk }^{ sd } -1,
\end{equation}
\centerline{$\forall s,d\in \mathcal{N}, \forall m,n,k \in \mathcal{N}: m \ne n \ne k$}.
Constraints (\ref{CwCalc1}, \ref{CwCalc2}, \ref{CwCalc3}) calculate the variable $c_{nmk}^{sd}$ for each node and different traffic flows, the three constraints are equivalent to the $c_{nmk }^{ sd } = b_{ nm }^{ sd }  b_{ mk }^{ sd }$, and they are used to maintain the linearity of the model that is otherwise lost due to the multiplication of variables. 

\begin{equation} \label{NxpCalc1}
X^{m}_{nk} =\sum _{ s \in \mathcal{N} } \sum _{ \substack{d\in \mathcal{N}\\ s\neq d }}{   max(\frac{c_{ nmk }^{ sd } \lambda ^{ sd },c_{ nmk }^{ ds } \lambda^{ds}}{B} )  },
\end{equation}
\centerline{$\forall m,n,k \in \mathcal{N}: m \ne n \ne k$}.
Constraint (\ref{NxpCalc1}) calculates the total number of coded ports at node $m$ that encodes the bidirectional traffic between nodes pair $(n,k)$. Note that the number of NC ports at a given node is determined according to the maximum flow of the bidirectional traffic demand of which the node is intermediate due to zero padding of the smaller flow to match the size of the larger flow. The presence of the $max()$ function in equation (\ref{NxpCalc1}) makes the model nonlinear, hence we use instead the following six constraints collectively to maintain the linearity of the model. 
\begin{equation} \label{NxpCalc}
w_{ nk }^{ m }=\sum _{ s \in \mathcal{N} } \sum _{ \substack{d\in \mathcal{N}\\ s\neq d }}{  c_{ nmk }^{ sd } \lambda ^{ sd }   }.
\end{equation}
 Constraint (\ref{NxpCalc}) calculates the traffic flow value from node $n$ to node $k$ passing through node $m$. To determine the maximum of the bidirectional flows (i.e. between $w^m_{nk}$ and $w^m_{kn}$), the variables $\Delta^{m+}_{nk}$ and $\Delta^{m+}_{nk}$ are introduced as binary variables, used to tell if the difference of opposite flows is positive or negative.
 The variable $\Delta^{m}_{nk}$ calculates the difference of the bidirectional traffic between nodes $n,k$ that passes through node $m$, calculated in Constraint (\ref{NxpCalc2}).  
\begin{equation} \label{NxpCalc2}
\Delta^m_{nk}=w_{ nk }^{ m }-w_{kn}^{m},
\end{equation}
\centerline{$\forall m,n,k \in \mathcal{N}: m \ne n \ne k$}.
\begin{equation} \label{NxpCalc3}
\Delta^m_{nk} \le M \Delta^{m+}_{nk}.
\end{equation}

\begin{equation} \label{NxpCalc4}
\Delta^m_{nk} \ge -M \Delta^{m-}_{nk}.
\end{equation}
The binary indicator variables $\Delta^{m+}_{nk}$ and $\Delta^{m-}_{nk}$ are used to tell if the bidirectional flow difference variable (i.e. $\Delta^{m}_{nk}$) is positive or negative. If $\Delta^{m}_{nk}$ is positive, then $\Delta^{m+}_{nk}$ is set to $1$, and if $\Delta^{m}_{nk}$ is negative, then $\Delta^{m-}_{nk}$ is set to $1$. The case of $\Delta^{m}_{nk}=0$ means both $\Delta^{m+}_{nk}$ and $\Delta^{m-}_{nk}$ can take the value $0$ or $1$, and to resolve this ambiguity, constraint (\ref{NxpCalc5}) is used to ensure that they are not both set to $1$. 
\begin{equation} \label{NxpCalc5}
\Delta^{m+}_{nk}+\Delta^{m-}_{nk} \le 1.
\end{equation}

\begin{equation} 
\label{NxpCalc6}
 X^{m}_{nk} = \left\{ 
  \begin{array}{l l}
     \frac{w_{ nk }^{ m }}{B} & \quad \text{if $\Delta^{m+}_{nk}$ =1}\\
    \frac{w_{kn}^{ m }}{B} & \quad \text{otherwise}
  \end{array} \right\}
\end{equation}

\centerline{$\forall m,s,d \in \mathcal{N},n,k \in \mathcal{N}_{ m }, n<k$}.
\vspace{5mm}
\begin{equation} \label{CalcNumXor}
X_{ m }=\sum _{ n\in \mathcal{N}_m } \sum _{\substack{ k\in \mathcal{N}_m\\ n<k }} X^{m}_{nk}
\end{equation}
Constraint (\ref{CalcNumXor}) calculates the  number of NC ports at each given node.

\section{Network Performance Evaluation}
\begin{figure}[h]
\centering
\begin{tikzpicture}[>=stealth',semithick]
    \tikzstyle{obj}  = [circle, minimum width=15pt, draw, inner sep=1pt, font=\scriptsize]
	\tikzstyle{obj2}=[ midway, above, fill=white, font=\scriptsize]
    \tikzstyle{every label}=[font=\scriptsize]

    	\node[obj] (1) at (0,0) {1};
	\node[obj] (2) at (1,1) {2};
    	\node[obj] (3) at (1,-1) {3};
     	\node[obj] (4) at (2,0.2) {4};
	 \node[obj] (5) at (3,0) {5};
	\node[obj] (6) at (3,-1.5) {6};
    	\node[obj] (7) at (4.4,0.2) {7};
     	\node[obj] (8) at (5.5,0.8) {8};
	\node[obj] (9) at (6.5,0.2) {9};
	\node[obj] (10) at (4.7,-2) {10};
    	\node[obj] (11) at (7.5,2) {11};
     	\node[obj] (12) at (8.2,0.2) {12};
	\node[obj] (13) at (7,-1.9) {13};
	\node[obj] (14) at (8.2,-1.2) {14};

    \path[-]   (1)    edge                node[obj2]  {260}       (2);
    \path[-]   (1)    edge                node[obj2]  {252}       (3);
    \path[-]   (1)    edge                node[obj2]   {324}       (4);
    \path[-]   (2)    edge                node[obj2]  {380}       (3);
    \path[-]   (2)    edge                node[obj2]  {868}       (7);
    \path[-]   (3)    edge                node[obj2]   {416}       (6);
     \path[-]   (4)    edge               node[obj2]  {1140}       (11);
    \path[-]   (4)    edge                node[obj2]   {248}       (5);
    \path[-]   (5)    edge                node[obj2]   {272}       (6);
     \path[-]   (5)    edge                node[obj2, below]  {292}       (7);
    \path[-]   (6)    edge                node[obj2, below]   {704}       (10);
    \path[-]   (6)    edge                node[obj2]  {1036}       (13);
     \path[-]   (7)    edge               node[obj2]  {212}       (8);
    \path[-]   (8)    edge                node[obj2]   {224}       (9);
    \path[-]   (9)    edge                node[obj2]   {752}       (10);
     \path[-]   (9)    edge               node[obj2]   {668}       (14);
    \path[-]   (9)    edge                node[obj2]  {536}       (12);
    \path[-]   (11)    edge                node[obj2]   {408}       (12);
    \path[-]   (11)    edge                node[obj2]   {684}       (14);
    \path[-]   (12)    edge                node[obj2]   {664}       (13);
    \path[-]   (13)    edge                node[obj2, below]   {353}       (14);
        \draw [dashed] (6,2) -- (6,-3);
     \draw [dashed] (3.6,2) -- (3.6,-3);
	 \draw [dashed] (1.5,2) -- (1.5,-3);

  \node[rectangle, text width=2cm, align=center] (n1) at (0.5,-2.8) {\small Pacific standard time (PST)};
    \node[rectangle, text width=2cm, align=center] (n1) at (2.5,-2.8) {\small Mountain standard time (MST)};
      \node[rectangle, text width=2cm, align=center] (n1) at (4.8,-2.8) {\small Central standard time (CST)};
        \node[rectangle, text width=2cm, align=center] (n1) at (7.2,-2.8) {\small Eastern standard time (EST)};
\end{tikzpicture}
\caption{The NSFNET topology}
\label{fig:NSFNET}
\end{figure}
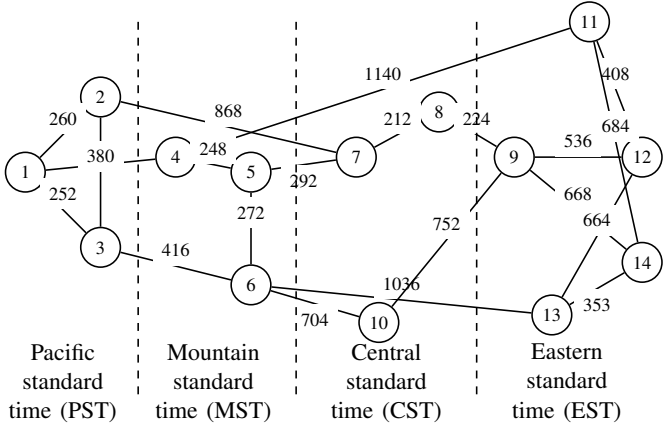

\begin{figure}[h]
\centering
\begin{tikzpicture}[>=stealth',semithick, scale=1.2]
    \tikzstyle{obj}  = [circle, minimum width=10pt, draw, inner sep=1pt, font=\scriptsize]
	\tikzstyle{obj2}=[ midway, above, fill=white, font=\scriptsize]
    \tikzstyle{every label}=[font=\scriptsize]
    	\node[obj] (1) at (0.3,2) {1};
	\node[obj] (2) at (0.1,1) {2};
    	\node[obj] (3) at (0,0) {3};
     	\node[obj] (4) at (0.5,-1) {4};
	 \node[obj] (5) at (0.1,-1.8) {5};
	 
	\node[obj] (6) at (1.5,1.2) {6};
    	\node[obj] (7) at (1.5,-0.2) {7};
     	\node[obj] (8) at (1.5,-2) {8};
	
	\node[obj] (9) at (2.8,0) {9};
	\node[obj] (10) at (2.8,-1.6) {10};
	
    	\node[obj] (11) at (3.8,1.2) {11};
     	\node[obj] (12) at (4,0) {12};
	\node[obj] (13) at (3.8,-1) {13};
	\node[obj] (14) at (3.8,-2) {14};
	
	\node[obj] (15) at (5.3,1) {15};
	\node[obj] (16) at (5,0) {16};
    	\node[obj] (17) at (5.2,-1.2) {17};
     	\node[obj] (18) at (5,-2.2) {18};
	
	 \node[obj] (19) at (6.5,2.2) {19};
	\node[obj] (20) at (6.3,1) {20};
    	\node[obj] (21) at (6.3,0.2) {21};
     	\node[obj] (22) at (6,-0.6) {22};
	\node[obj] (23) at (6.2,-1.3) {23};
	\node[obj] (24) at (6.1,-2) {24};

    \path[-]   (1)    edge                node[obj2]  {252}       (2);
    \path[-]   (2)    edge                node[obj2]  {216}       (3);
    \path[-]   (3)    edge                node[obj2]   {}       (4);
    \path[-]   (4)    edge                node[obj2, below]  {220}       (5);
    \path[-]   (3)    edge                node[obj2]  {432}       (5);
    \path[-]   (2)    edge                node[obj2, below]   {360}       (6);
     \path[-]   (3)    edge               node[obj2]  {304}       (7);
    \path[-]   (4)    edge                node[obj2]   {280}       (7);
    \path[-]   (5)    edge                node[obj2]   {368}       (8);
     \path[-]   (1)    edge                node[obj2, below]  {364}       (6);
    \path[-]   (7)    edge                node[obj2, below]   {464}       (8);
    \path[-]   (6)    edge                node[obj2]  {262}       (7);
     \path[-]   (6)    edge               node[obj2]  {360}       (9);
    \path[-]   (7)    edge                node[obj2]   {328}       (9);
    \path[-]   (9)    edge                node[obj2]   {440}       (10);
     \path[-]   (8)    edge               node[obj2]   {272}       (10);
    \path[-]   (9)    edge                node[obj2]  {364}       (11);
    \path[-]   (6)    edge                node[obj2]   {572}       (11);
    \path[-]   (11)    edge                node[obj2]   {288}       (12);
    \path[-]   (12)    edge                node[obj2]   {236}       (13);
    \path[-]   (13)    edge                node[obj2, below]   {244}       (14);
    
     \path[-]   (9)    edge                node[obj2]  {320}       (12);
    \path[-]   (10)    edge                node[obj2]  {320}       (13);
    \path[-]   (10)    edge                node[obj2]   {268}       (14);

    \path[-]   (11)    edge                node[obj2, below]  {344}       (15);
    \path[-]   (11)    edge                node[obj2]  {648}       (19);
    \path[-]   (12)    edge                node[obj2]   {280}       (16);
     \path[-]   (13)    edge               node[obj2]  {332}       (17);
    \path[-]   (14)    edge                node[obj2]   {368}       (18);
    \path[-]   (15)    edge                node[obj2]   {}       (16);
     \path[-]   (15)    edge                node[obj2, below]  {272}       (20);
    \path[-]   (16)    edge                node[obj2, below]   {224}       (17);
    \path[-]   (16)    edge                node[obj2]  {288}       (21);
     \path[-]   (16)    edge               node[obj2]  {272}       (22);
    \path[-]   (17)    edge                node[obj2]   {280}       (18);
    \path[-]   (17)    edge                node[obj2]   {168}       (22);
     \path[-]   (17)    edge               node[obj2]   {364}       (23);
    \path[-]   (18)    edge                node[obj2]  {280}       (24);
    \path[-]   (19)    edge                node[obj2]   {188}       (20);
    \path[-]   (20)    edge                node[obj2]   {216}       (21);
    \path[-]   (21)    edge                node[obj2]   {164}       (22);
    \path[-]   (22)    edge                node[obj2, rotate=-90]   {260}       (23);
    \path[-]   (23)    edge                node[obj2, rotate=90]   {180}       (24);
    \draw [dashed,smooth] (4,2) --(4.5,0)-- (5.5,-1)--(5.7,-3);
     \draw [dashed] (2.6,2) -- (2.6,-3);
	 \draw [dashed] (1,2) -- (1,-3);
	   \node[rectangle] (n1) at (0.5,-2.5) {PST};
    \node[rectangle] (n1) at (2,-2.5) {MST};
      \node[rectangle] (n1) at (4,-2.5) {CST};
        \node[rectangle] (n1) at (6,-2.5) {EST};
\end{tikzpicture}
\caption{The USNET topology}
\label{fig:USNET}
\end{figure}
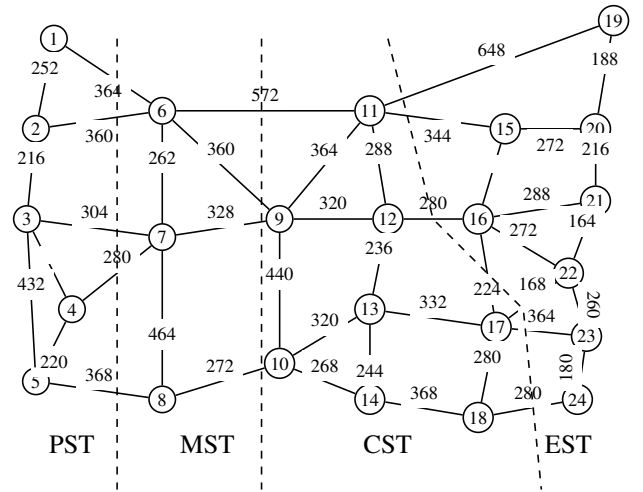

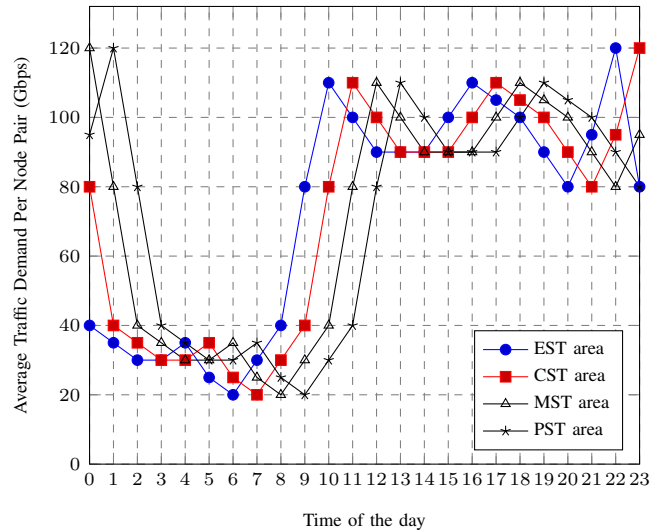
\begin{figure}
\centering
 	\begin{tikzpicture}
		\begin{axis} [width=\linewidth,font=\scriptsize, grid=both,ylabel near ticks,ylabel={Average Traffic Demand Per Node Pair (Gbps)}, xlabel={Time of the day},xmin=0, xmax=23, ymin=0, xtick={0,1,2,3,4,5,6,7,8,9,10,11,12,13,14,15,16,17,18,19,20,21,22,23},legend style={nodes=right},legend pos= south east]]
			\addplot coordinates{(0,40)(1,35)(2,30)(3,30)(4,35)(5,25)(6,20)(7,30)(8,40)(9,80)(10,110)(11,100)(12,90)(13,90)(14,90)(15,100)(16,110)(17,105)(18,100)(19,90)(20,80)(21,95)(22,120)(23,80)};
			\addlegendentry{EST area}
			\addplot coordinates{(0,80)(1,40)(2,35)(3,30)(4,30)(5,35)(6,25)(7,20)(8,30)(9,40)(10,80)(11,110)(12,100)(13,90)(14,90)(15,90)(16,100)(17,110)(18,105)(19,100)(20,90)(21,80)(22,95)(23,120)};
			\addlegendentry{CST area}
			\addplot [mark =triangle]coordinates{(0,120)(1,80)(2,40)(3,35)(4,30)(5,30)(6,35)(7,25)(8,20)(9,30)(10,40)(11,80)(12,110)(13,100)(14,90)(15,90)(16,90)(17,100)(18,110)(19,105)(20,100)(21,90)(22,80)(23,95)};
			\addlegendentry{MST area}
			\addplot coordinates{(0,95)(1,120)(2,80)(3,40)(4,35)(5,30)(6,30)(7,35)(8,25)(9,20)(10,30)(11,40)(12,80)(13,110)(14,100)(15,90)(16,90)(17,90)(18,100)(19,110)(20,105)(21,100)(22,90)(23,80)};
			\addlegendentry{PST area}
		\end{axis} 
	\end{tikzpicture}
\caption{Average traffic demands at different times of the day[4]}
\label{fig:Traffic}
\end{figure}

We use regular and common network topologies in our analysis. For regular topologies we analyze the ring, line, star and full mesh topologies and  for real world core networks we consider the NSFNET and the USNET networks topologies in USA. In Fig. \ref{fig:NSFNET} the NSFNET is shown and is composed of 14 nodes and 21 bidirectional links, with an average hop count equal to 2.17. The USNET, depicted in Fig. \ref{fig:USNET}, has an average hop count of 3, with 43 links connecting 24 nodes. 
\\We model traffic demands by using the average traffic demands of the network as shown in Fig. \ref{fig:Traffic} \cite{Dong2011}. The traffic follows a uniform probability distribution with values ranging from 20 Gb/s to 120 Gb/s. The peak value is observed at 22:00. This peak shifts according to the time zones by 1 hour. The actual traffic matrix is generated randomly and ensures that the average traffic at each time zone (including intra-zone traffic and inter-zone traffic) is that of Fig. \ref{fig:Traffic}.  This traffic model shows two kinds of variations, the first is the variation between different times of the day and the second between different nodes at a given time point. The traffic demands between node pairs takes values between 10 and 230 Gbps. 



\begin{center}
\begin{table}[h]
 \small
 \centering 
  \caption{Network Parameters} 

 \begin{tabular}{l l} 
 \hline\hline 
 Parameter & Value \\ [0.5ex] 
 \hline 
 Separating distance between EDFAs & 80 km  \\ 
 Number of wavelengths in a single fibre (W) & 16  \\
 Capacity of a single wavelength (B) & 40 Gbps \\
 Conventional port power consumption (Pp) \cite{CRS1} & 1 kW  \\
 NC port power consumption (Px) & 1.1 kW  \\ 
 Transponder power consumption (Pt) \cite{transponder} & 73 W  \\
Optical Switch power consumption (PO) \cite{Glimmerglass}&	85 W\\
 MUX/DeMUX power consumption \cite{MUX}	&16 W\\
EDFAs power consumption  (Pe)	\cite{EDFA}& 8 W\\
 \hline 
 \end{tabular}
 \label{table:parameters}
 \end{table}
\end{center}
Other network parameters are shown in Table \ref{table:parameters} , including the power consumption of various network elements. 
The power consumption of the network devices is derived from \cite{CRS1,transponder,Glimmerglass, MUX,EDFA}. The 100 Watts increase in power consumption for the NC ports represents a moderate estimate of the power consumption of the processing of the XOR operation and the added transmitter and amplifier at the physical interface module of the port. In the calculations, a partially used wavelength was assumed to consume part of a router port and a transponder proportional to traffic volume. This assumes either grooming or a proportional power profile for router ports and transponders and such a proportional power profile is desirable and is a goal of the current equipment manufactures and grooming can be practically achieved. 
The optimization was carried out using IBM CPLEX software suite and the AMPL language and was executed on a high performance computing cluster with 256GB RAM and 16 cores of CPU.  The problem  falls under the NP-complete complexity class as it can be reduced to the standard multi-commodity flow problem (an NP-complete problem), and is further complicated by the need to consider the assignment of  coding nodes and accounting for the zero padding and partitioning of traffic. \\
For the solution scenarios considered in the paper, the MILP model runs in the order of days. The heuristic, on the other hand, have a running time in the order of seconds.  
\subsection{Example Topologies}
Fig. \ref{fig:NSFNET_Power} and Fig. \ref{fig:USNET_Power} show the power consumption of the NSFNET and USNET topologies, on a full day at 2 hours granularity, respectively. Daily average savings of 27\% and 33\% are observed as a result of using network coding in the NSFNET and USNET topologies, respectively. Note that the higher average hop count of the USNET topology has a limited effect in terms of increasing the power savings. Although a topology with a higher hop count means higher number of conventional ports are replaced by Network Coding enabled ports, however it also means more conventional ports are required to establish flows between neighbouring nodes, in which scenario the data  flows cannot be encoded. Therefore, the total power savings produced by using network coding depends on the ratio of the traffic between neighbouring nodes (which cannot be coded) to the traffic between non-neighbouring nodes (which can be coded). Higher savings can be obtained for networks of high hop counts as discussed below.

\begin{figure}
\centering
 \begin{tikzpicture}[scale=1]
\begin{axis} [
 xtick=data, grid=both,
ylabel near ticks,
ylabel={number of Ports},
xlabel={Node number},
ybar=1.5pt,
bar width =5pt,  font=\scriptsize
]
\addplot[fill=red!50]
coordinates{(1,9.15)(2,10.5)(3,14.2)(4,22.4)(5,21.6)(6,38.5)(7,24)(8,15.7)(9,25.54)(10,4.9)(11,13.3)(12,8.3)(13,14.4)(14,6.7)};
\addlegendentry{NC ports}
\addplot[fill=blue!10] coordinates{(1,27)(2,29.5)(3,26)(4,26)(5,26)(6,30)(7,28)(8,25)(9,26.5)(10,25)(11,27)(12,26.5)(13,27)(14,26)};
\addlegendentry{Conventional ports}
\end{axis} 
\end{tikzpicture}
\caption{Comparison between the number of the two port types in the NSFNET nodes}
\label{fig:Port_dist}
\end{figure}
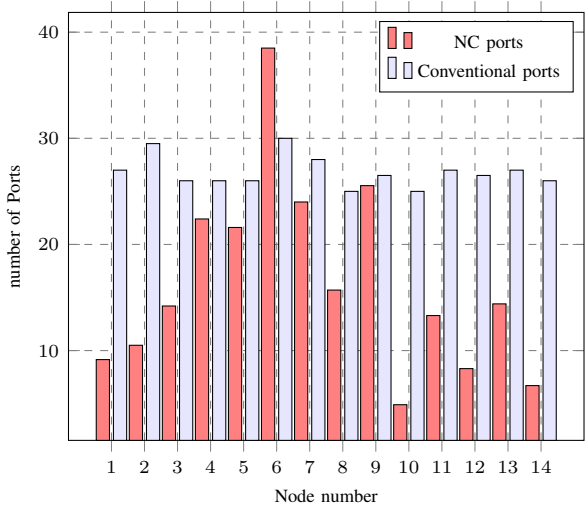

Fig. \ref{fig:Port_dist} compares the distribution of conventional ports and NC ports across the NSFNET nodes. The conventional ports tend to be almost equally distributed among nodes, while NC ports are used more in the middle of the network at nodes with high nodal degree. This is because middle nodes are more likely to serve as intermediate nodes to traffic flows. Node 6 deploys the highest number of NC ports as it is located in the network centre, having a nodal degree equal to 4. Nodes of lower hop count and/or not centrally located deploy lower number of NC ports.

\begin{figure}
\centering
 	\begin{tikzpicture}[scale=1]
		\begin{axis} [grid=both,ylabel near ticks,ylabel={Power (W)}, xlabel={Time of the day},xmin=0, xmax=22, ymin=0, xtick={0,2,4,6,8,10,12,14,16,18,20,22},legend style={nodes=right},legend pos= north west ,  font=\scriptsize]
			\addplot coordinates{(0,745500)(2,424800)(4,274700)(6,237800)(8,255600)(10,617300)(12,779300)(14,792800)(16,830900)(18,876600)(20,787900)(22,900300)};
			\addlegendentry{NC (MILP) }
			\addplot coordinates{(0,843462)(2,492409)(4,299505)(6,262153)(8,283473)(10,725495)(12,870417)(14,888594)(16,919483)(18,987829)(20,872051)(22,985507)};
			\addlegendentry{NC (Heuristic)}
			\addplot[mark=triangle] coordinates{(0,1042262)(2,608149)(4,366104)(6,319863)(8,354503)(10,892867)(12,1041468)(14,1092694)(16,1130832)(18,1202788)(20,1076467)(22,1222775)};
			\addlegendentry{Conventional}
		\end{axis} 
	\begin{axis}[
  axis y line*=right,
  axis x line=none,
  ymin=10, ymax=100,
  xmin=0, xmax=22,
  ylabel=$saving\%$,
    ylabel near ticks, 
    legend style={at={(0.03,0.65)},anchor=west, font=\scriptsize}
]
\addplot[dashed,mark=x,red, thick] 
  coordinates{
    (0,19.07)(2,19.03)(4,18.2)(6,18.04)(8,20.03)(10,18.74)(12,16.42)(14,18.67)(16,18.68)(18,17.87)(20,18.9)(22,19.4) 
}; 
 \addlegendentry{Heuristic savings}
\addplot[dashed,mark=triangle,blue, thick] 
  coordinates{
    (0,28.47)(2,30.15)(4,25)(6,25.66)(8,27.9)(10,30.86)(12,25.2)(14,27.4)(16,26.5)(18,27.1)(20,26.8)(22,26.4) 
}; 
\addlegendentry{MILP savings}
\end{axis}
	\end{tikzpicture}
\caption{The NSFNET network power consumption with and without network coding}
\label{fig:NSFNET_Power}
\end{figure}
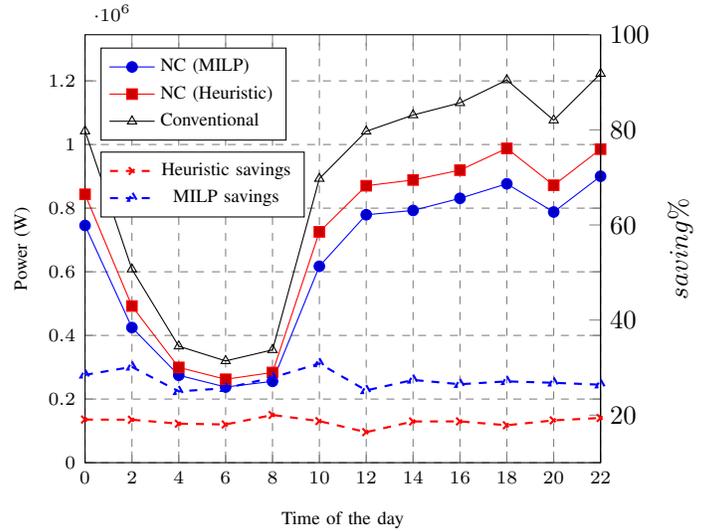

\begin{figure}
\centering
 	\begin{tikzpicture}[scale=1]
		\begin{axis} [grid=both,ylabel near ticks,ylabel=Power (W),xlabel={Time of the day}, xmin=0, xmax=20, ymin=0, xtick={0,2,4,6,8,10,12,14,16,18,20},legend style={nodes=right},legend pos= north west,  font=\scriptsize]  
			\addplot coordinates{(0,2819378)(4,1060265)(8,1042324)(12,3109619)(16,3432625)(20,3111130)};
			\addlegendentry{NC (MILP)}
			\addplot coordinates{(0,3278347)(4,1226000)(8,1213397)(12,3616314)(16,4044564)(20,3630571)};
			\addlegendentry{NC (Heuristic) }
			\addplot[mark=triangle] coordinates{(0,4163066)(4,1586099)(8,1582017)(12,4603695)(16,5110054)(20,4631329)};
			\addlegendentry{Conventional}
		\end{axis} 
\begin{axis}[
  ymin=10, ymax=100,
  xmin=0, xmax=20,
  ylabel={$saving\%$},
  ylabel near ticks, 
  axis y line*=right, 
  hide x axis, 
  legend style={at={(0.03,0.65)},anchor=west, font=\scriptsize}
]
\addplot[dashed,mark=x,red, thick] 
  coordinates{
    (0,21.3)(4,22.7)(8,23.3)(12,21.4)(16,20.85)(20,21.6)
}; \addlegendentry{Heuristic savings}
\addplot[dashed,mark=triangle,blue, thick] 
  coordinates{
    (0,32.3)(4,33.15)(8,34.1)(12,32.45)(16,32.8)(20,32.8)
}; 
\addlegendentry{MILP savings}
\end{axis}
	\end{tikzpicture}
\caption{The USNET network power consumption with and without network coding}
\label{fig:USNET_Power}
\end{figure}
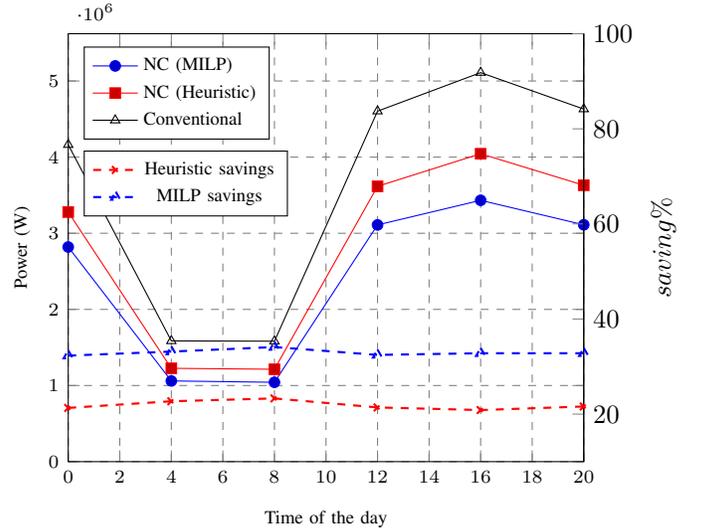

The routes selected by the solution of the MILP model show no departure from the minimum hop routes selected by the conventional scenario. 
The energy efficient routing heuristic used in our case can be simply described by routing all demands following the minimum hop path between the source and destination nodes, selecting the shortest path in case two alternative paths have the same minimum hops. Based on this flow allocation, network coding takes place at all intermediate nodes, then the total power consumption and the amount of conventional ports and NC ports are calculated. This fact can help facilitate the migration to the new architecture. Consequently, the amount of required coded ports at each node can be estimated at the time of design. \\
Fig. \ref{fig:NSFNET_Power} and \ref{fig:USNET_Power} show that the average power savings of the energy efficient heuristic in network coding enabled networks (19\% and 22\% considering the NSFNET and the USNET networks, respectively) compared with the energy efficient network coding MILP model. \\

\subsection{Regular Topologies}
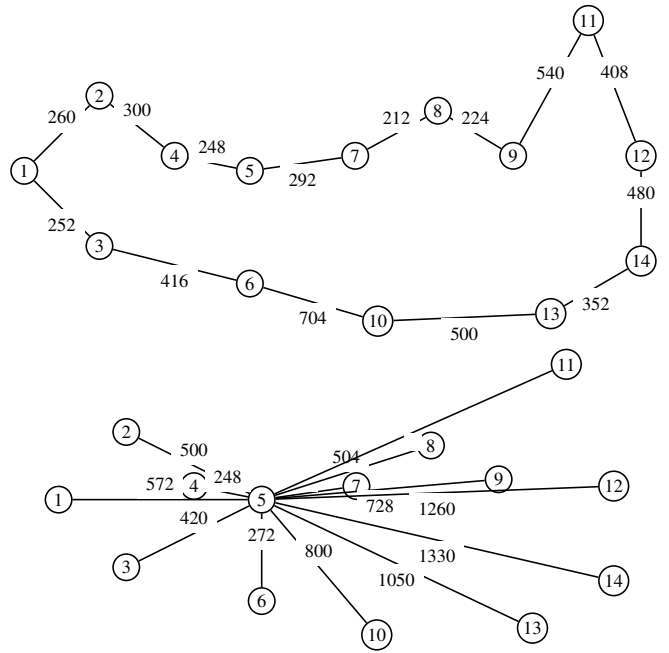
\begin{figure}[h]
\centering
\begin{tikzpicture}[>=stealth',semithick, scale=1]
    \tikzstyle{obj}  = [circle, minimum width=10pt, draw, inner sep=1pt, font=\scriptsize]
	\tikzstyle{obj2}=[ midway, above, fill=white, font=\scriptsize]
    \tikzstyle{every label}=[font=\scriptsize]

    	\node[obj] (1) at (0,0) {1};
	\node[obj] (2) at (1,1) {2};
    	\node[obj] (3) at (1,-1) {3};
     	\node[obj] (4) at (2,0.2) {4};
	 \node[obj] (5) at (3,0) {5};
	\node[obj] (6) at (3,-1.5) {6};
    	\node[obj] (7) at (4.4,0.2) {7};
     	\node[obj] (8) at (5.5,0.8) {8};
	\node[obj] (9) at (6.5,0.2) {9};
	\node[obj] (10) at (4.7,-2) {10};
    	\node[obj] (11) at (7.5,2) {11};
     	\node[obj] (12) at (8.2,0.2) {12};
	\node[obj] (13) at (7,-1.9) {13};
	\node[obj] (14) at (8.2,-1.2) {14};

    \path[-]   (1)    edge                node[obj2]  {260}       (2);
    \path[-]   (2)    edge                node[obj2]  {300}       (4);
     \path[-]   (4)    edge               node[obj2]  {248}       (5);
     \path[-]   (5)    edge                node[obj2, below]  {292}       (7);
     \path[-]   (7)    edge               node[obj2]  {212}       (8);
    \path[-]   (8)    edge                node[obj2]   {224}       (9);
    \path[-]   (9)    edge                node[obj2]  {540}       (11);
    \path[-]   (11)    edge                node[obj2]   {408}       (12);
    \path[-]   (12)    edge                node[obj2]   {480}       (14);
    \path[-]   (14)    edge                node[obj2, below]   {352}       (13);
     \path[-]   (13)    edge                node[obj2, below]   {500}       (10);
    \path[-]   (10)    edge                node[obj2, below]   {704}       (6);
    \path[-]   (6)    edge                node[obj2, below]   {416}       (3);
     \path[-]   (3)    edge                node[obj2, below]   {252}       (1);
\end{tikzpicture}

\begin{tikzpicture}[>=stealth',semithick, scale=0.9]
    \tikzstyle{obj}  = [circle, minimum width=10pt, draw, inner sep=1pt, font=\scriptsize]
	\tikzstyle{obj2}=[ midway, above, fill=white, font=\scriptsize]
    \tikzstyle{every label}=[font=\scriptsize]
    	\node[obj] (1) at (0,0) {1};
	\node[obj] (2) at (1,1) {2};
    	\node[obj] (3) at (1,-1) {3};
     	\node[obj] (4) at (2,0.2) {4};
	 \node[obj] (5) at (3,0) {5};
	\node[obj] (6) at (3,-1.5) {6};
    	\node[obj] (7) at (4.4,0.2) {7};
     	\node[obj] (8) at (5.5,0.8) {8};
	\node[obj] (9) at (6.5,0.3) {9};
	\node[obj] (10) at (4.7,-2) {10};
    	\node[obj] (11) at (7.5,2) {11};
     	\node[obj] (12) at (8.2,0.2) {12};
	\node[obj] (13) at (7,-1.9) {13};
	\node[obj] (14) at (8.2,-1.2) {14};
    \path[-]   (1)    edge                node[obj2]  {572}       (5);
    \path[-]   (2)    edge                node[obj2]  {500}       (5);
     \path[-]   (3)    edge               node[obj2]  {420}       (5);
     \path[-]   (4)    edge                node[obj2]  {248}       (5);
     \path[-]   (6)    edge               node[obj2]  {272}       (5);
    \path[-]   (7)    edge                node[obj2]   {}       (5);
    \path[-]   (8)    edge                node[obj2]  {504}       (5);
    \path[-]   (9)    edge                node[obj2, below]   {728}       (5);
    \path[-]   (10)    edge                node[obj2]   {800}       (5);
    \path[-]   (11)    edge                node[obj2, below]   {}(5);
     \path[-]   (12)    edge                node[obj2, below]   {1260}       (5);
    \path[-]   (13)    edge                node[obj2, below]   {1050}       (5);
    \path[-]   (14)    edge                node[obj2, below]   {1330}       (5);
\end{tikzpicture}
\caption{The NSFNET connected by ring and star topologies}
\label{fig:NSFNET_RingStar}
\end{figure}

To investigate the impact of the topology, we reconfigured the NSFNET network into a bidirectional ring, and a star as seen in Fig. \ref{fig:NSFNET_RingStar}, as well as   a line (by removing link (12-14) from the ring), and a full mesh. Note that a star topology can result in practice if the traffic due to a large data centre dominates the network. The full mesh can be an attractive network topology if delay and power consumption are key metrics \cite{Dong2012}. The ring and line topologies have high average hop counts and hence were considered. The distances (in km) shown on these links are estimates according to nodes locations.  Network coding contributes the highest reduction in network power consumption (33\% daily average saving) in the line topology. The high power saving of the line topology is attributed to its high average hop count of 5, increasing the number of intermediate nodes, which increases the number conventional ports to be replaced by NC ports. On the other hand network coding adds no gain to the full mesh topology, as all the demands are routed through single hop routes. The savings obtained by the other topologies exist between these two extremes. Implementing network coding in a bidirectional ring topology (average hop count of 3.77) has saved 30\% of the network power consumption while a star topology centred at node 5 with an average hop count of 1.85 has saved 16\% of the network power consumption.
The topologies we considered, vary from the full mesh topology having the maximum number of links, to the star topology, having the lowest number of links while maintaining connectivity. 


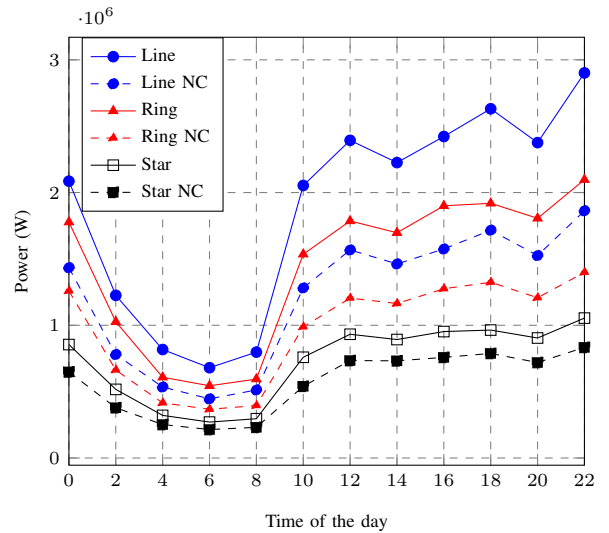
\begin{figure}[h]
\centering
\begin{tikzpicture}[scale=1]
\begin{axis} [
xtick=data, grid=both,xmin=0,xmax=22,
ylabel={Power (W)},
xlabel={Time of the day},
ylabel near ticks,
legend style={at ={(0.3,1)}}, 
legend cell align=left,  font=\scriptsize
]
\addplot[color=blue,mark=*]
coordinates{
(0,2085600.7)
(2,1225577.4)
(4,817305)
(6,680055.7)
(8,797786.7)
(10,2053211.4)
(12,2393392)
(14,2225690)
(16,2422230.6)
(18,2631259.2)
(20,2376232)
(22,2901409)
};
\addlegendentry{Line}
\addplot[color=blue,style=dashed,mark=*]
coordinates{
(0,1434247)
(2,779776)
(4,534785)
(6,445775.7)
(8,512737.6)
(10,1281282)
(12,1567477)
(14,1462270)
(16,1574816)
(18,1716464)
(20,1525498.4)
(22,1862719.6)};
\addlegendentry{Line NC}
\addplot[color=red,mark=triangle*] 
coordinates{
(0,1777364)
(2,1026427)
(4,608469)
(6,544117)
(8,594776)
(10,1535106)
(12,1786038)
(14,1697315)
(16,1900082)
(18,1919026)
(20,1806029)
(22,2097044)
};
\addlegendentry{Ring}
\addplot[color=red,style=dashed,mark=triangle*]
coordinates{
(0,1259302.8875)
(2,662782)
(4,415219.6)
(6,367133)
(8,395174)
(10,987730)
(12,1204800)
(14,1164140)
(16,1275370)
(18,1324288)
(20,1207564)
(22,1398718)
};
\addlegendentry{Ring NC}
\addplot[color=black,mark=square]
coordinates
{
(0,854806)
(2,517060)
(4,321184)
(6,270637)
(8,295880)
(10,759070)
(12,932407)
(14,892617)
(16,952658)
(18,963972)
(20,904110)
(22,1054223)
};
\addlegendentry{Star}
\addplot[color=black,style=dashed,mark=square*]
coordinates
{(0,648749)(2,378902)(4,252120)(6,214441.6)(8,231076)(10,540262)(12,734151.5)(14,732234.6)(16,758088)(18,787826.5)(20,719435.7)(22,833386)};
\addlegendentry{Star NC}
\end{axis} 
\end{tikzpicture}
\caption{Ring and Star topologies power consumption with and without network coding (MILP results)}
\label{RingStarPower}
\end{figure}

\begin{figure}[h]
\centering
\begin{tikzpicture}[scale=1]
\begin{axis} [
,symbolic x coords={Line,Ring,NSFNET,Star, Full Mesh},
 xtick=data, grid=both,
ylabel near ticks,
ylabel={Savings \%},
xlabel={Topology},
font=\scriptsize
]
\addplot+[only marks] plot[error bars/.cd, y dir=both, y explicit]
coordinates{
        (Line,35) +- (2.6,3.87)
    (Ring,32.7) +- (2.96,3.55) 
    (NSFNET,27.3) +- (3.56,2.3)
    (Star,21.93)+-(6.9,3.96)
    (Full Mesh, 0)+-(0,0)
};
\end{axis} 
\end{tikzpicture}
\caption{Maximum, minimum and average daily power savings in different network coded topologies}
\label{diffTop}
\end{figure}
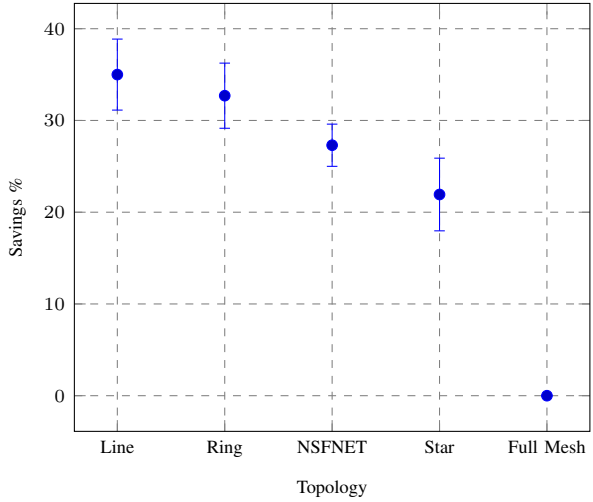
We extended the physical network topology MILP design model in \cite{Dong2012} to perform network coding at intermediate nodes of bidirectional traffic flows. We carried out extensive scenario evaluations. For example examined the optimum physical topology with a  constraint of a minimum nodal degree of 1 per node, and 21 links (number of links in the NSFNET). The optimum topology considering network coding was in agreement with that obtained in \cite{Dong2012}.

\section{Network coding with partitioning} \label{partitioning}
\begin{figure}[h]
  \centering
    \includegraphics[width=0.5\textwidth]{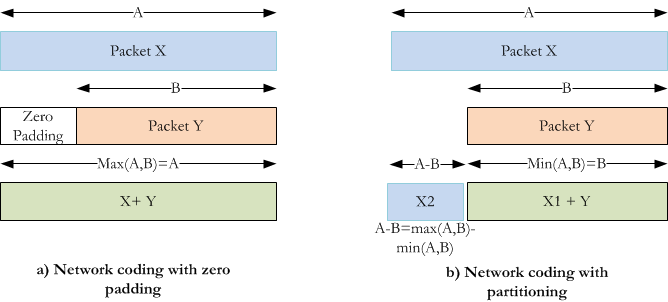}
    \caption{Approaches to ports calculation}
    \label{fig:ZP_paritioning}
\end{figure}

In the previous sections we assumed that the smaller of the two bidirectional flows (packets) is padded with zeros before encoding it with the other flow; hence the maximum of the two bidirectional flows is used to calculate the resources used. An alternative approach to prepare packets for network coding is to partition the bigger packet (flow) into two parts, the first part has the size of the smaller packet (flow) to be encoded, and the other part will be routed normally without encoding. Fig. \ref{fig:ZP_paritioning} (a) and (b) shows the zero padding approach (a) and the partitioning approach (b), respectively. 
The zero padding approach is the easier of the two, but less resource efficient. The partitioning approach, on the other hand, saves resources at the expense of adding control and management overheads to decide when and how to partition packets for the maximum benefit. 
We update the model in Section \ref{milp} to support the partitioning approach by changing the previous definition of the variable $Y_{nk}^m$ to represent the difference between the two packet sizes calculated as
\begin{equation} \label{NxpCalcPart}
Y_{nk}^m = \frac{w_{nk}^m - w_{nk}^m}{B},
\end{equation}
and changing the variable $X_{nk}^m$ to represent the minimum of the opposite flows, this will mean constraint (\ref{NxpCalc1}) will change into 
\begin{equation} \label{NxpCalcMin}
X^{m}_{nk} =\sum _{ s \in \mathcal{N} } \sum _{ \substack{d\in \mathcal{N}\\ s\neq d }}{   min(\frac{c_{ nmk }^{ sd } \lambda ^{ sd },c_{ nmk }^{ ds } \lambda^{ds}}{B} )}.
\end{equation}
and constraint (\ref{NxpCalc6}) will be replaced by

\begin{equation} \label{NxpCalc6Min}
 X^{m}_{nk} = \left\{ 
  \begin{array}{l l}
     \frac{w_{ nk }^{ m }}{B} & \quad \text{if $\Delta^{m-}_{nk}$ =1}\\
    \frac{w_{kn}^{ m }}{B} & \quad \text{otherwise}
  \end{array}. \right.
\end{equation}

\begin{figure}[h]
\centering
 	\begin{tikzpicture}[scale=1]
		\begin{axis} [width=\linewidth,font=\scriptsize, grid=both,ylabel near ticks,ylabel={Power (W)}, xlabel={Time of the day},xmin=0, xmax=22, ymin=0, xtick={0,2,4,6,8,10,12,14,16,18,20,22},legend style={nodes=right},legend pos= south east]]
			\addplot coordinates{(0,1042262)(2,608149)(4,366104)(6,319863)(8,354503)(10,892867)(12,1041468)(14,1092694)(16,1130832)(18,1202788)(20,1076467)(22,1222775)};
			\addlegendentry{Conventional}
			\addplot coordinates{(0, 745500)(2, 424800)(4, 274700)(6, 237800)(8, 255600)(10,617300)(12,779300)(14,792800)(16,830900)(18,876600)(20,787900)(22,900300)};
			\addlegendentry{NC (zero padding)}
			\addplot [mark =triangle]coordinates{(0, 733798)(2, 416678)(4, 271327)(6, 234901)(8,252318)(10,601619)(12,769686)(14,780420)(16,817159)(18,865244)(20,778044)(22,886841)};
			\addlegendentry{NC (partitioning)}
			\addplot coordinates{(0, 735005)(2, 417127)(4, 271593)(6, 235541)(8, 252105)(10,603411)(12,774953)(14,783252)(16,819928)(18,871149)(20,778144)(22,889256)};			\addlegendentry{NC Heuristic (partitioning)}
		\end{axis} 
	\end{tikzpicture}
\caption{Power consumption of the NSFNET under  NC with partitioning approach}
\label{fig:MaxMin}
\end{figure}
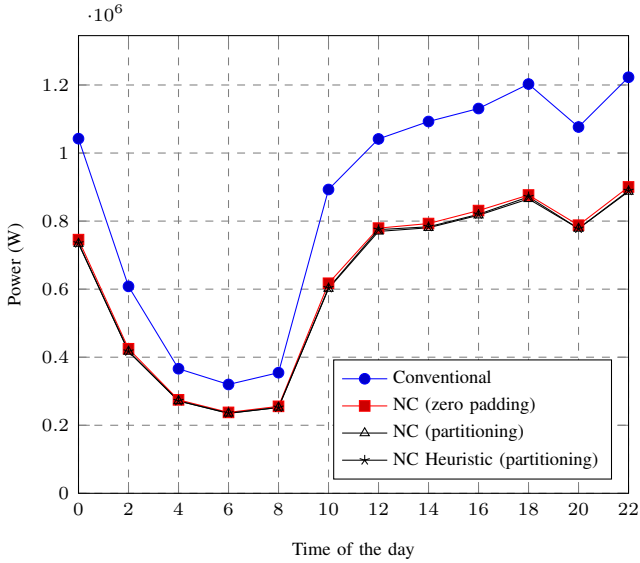

The network coding with partitioning is subject to all the constraints in Section III except for constraint (9) that calculates the total number of conventional ports which is replaced by
\begin{equation} \label{NxpCalcPart2}
Y_m = \sum_{n \in \mathcal{N}_m} \sum_{s \in \mathcal{N}} \sum_{\substack{d \in \mathcal{N} \\ s \ne d}}\frac{w^{sd}_{nm}}{B}+ \sum_{n \in \mathcal{N}_m} \sum_{\substack{k \in \mathcal{N}_m \\ k < n}}Y^m_{nk}
\end{equation}
The first term of equation (\ref{NxpCalcPart2}) is used to calculate the number of conventional ports in the zero padding case, where the zero padding case used the max() function in (\ref{NumPorts1}). The use of the max function was necessary as traffic was asymmetric. If source nodes, for instance, sum the traffic leaving plus entering and divide this sum by the wavelength rate, this will lead to an underestimation of the required number of ports. The correct number for the zero padding approach is dictated by the larger of the two traffic volumes leaving and entering the port, hence the max() function is needed in (\ref{NumPorts1}). In the partitioning case, however, the asymmetry is removed by the partitioning process and the residual traffic is handled using conventional network ports. \\
The first term of equation (\ref{NxpCalcPart2}) accounts for the conventional ports at source and destination nodes, while the second term accounts for the residual flow from the partitioning process. For example, if the bidirectional flow is 80Gbps in one direction and 50Gbps in the other, then in intermediate nodes 50Gbps will use a coded port while the rest (i.e. 30Gbps) will be served using conventional ports. \\
Fig. \ref{fig:MaxMin} shows that the improvement obtained by implementing the partitioning approach is limited compared to the zero padding approach under the uniform traffic profile. This is because the bidirectional flows of the traffic demands are comparable. The advantage of the partitioning approach under an asymmetric traffic profile will be discussed later.  

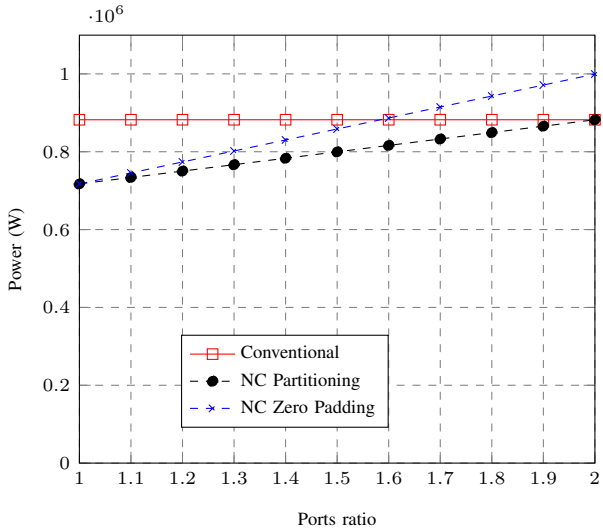
\begin{figure}[h]
\centering
\begin{tikzpicture}[scale=1]
\begin{axis} [
xtick=data, grid=both,xmin=1,xmax=2,ymin=0,xlabel={Ports ratio},
ylabel={Power (W)},
ylabel near ticks,
legend style={at ={(0.6,0.3)}}, 
legend cell align=left,  font=\scriptsize
]
	\addplot[domain=1:2, samples=11,color=red, mark=square] {882182};
	\addlegendentry{Conventional}

	\addplot[domain=1:2,samples=11,color=black,style=dashed, mark=*] coordinates
{
(1, 	717311)
(1.1 ,	733798)
(1.2 ,	750286)
(1.3 ,	766773)
(1.4,	783261)
(1.5,	799748)
(1.6,	816236)
(1.7,	832723)
(1.8,	849211)
(1.9,	865698)
(2,	882182)
};
	\addlegendentry{NC Partitioning}
	
		\addplot[domain=1:2,samples=11,color=blue,style=dashed, mark=x] coordinates
{
(1, 	717311)
(1.1 ,	745533)
(1.2 ,	773756)
(1.3 ,	801978)
(1.4,	830201)
(1.5,	858423)
(1.6,	886646)
(1.7,	914868)
(1.8,	943091)
(1.9,	971313)
(2,	 999536)
};
	\addlegendentry{NC Zero Padding}
\end{axis} 
\end{tikzpicture}
\caption{Power consumption versus ports ratio using the MILP model}
\label{fig:SensitivityR}
\end{figure}

All the previous results are based on the estimation of the NC port's power consumption given in Section \ref{milp}. Fig. \ref{fig:SensitivityR} studies how the savings obtained by network coding are effected as $r$ (the ratio of the network coding scheme's power consumption (i.e. router ports and transponders) and the conventional scheme) grows. The savings achieved compared to the conventional case reduces as $r$ increases. While the zero padding approach out performs the conventional approach up to $r=1.6$, the benefit of the partitioning approach can be observed up to $r=2$.

Fig. \ref{fig:SensitivityR} shows the power consumption dependence in the network coding case on the ports ratio $r=\frac{p_x+p_t}{p_p+p_t}$ for the conventional and the the network coding case with packet partitioning and zero padding under random traffic using the MILP model. It shows that the port ratio has an impact on the savings, which are better in the case of the packet partitioning. We observe savings as long as the power consumption of the NC port (including the transponder) is less than twice that of the power consumption of the conventional port.

\section{Network coding considering the bypass routing approach}
In the previous sections we showed that network coding can provide up to 33\% power savings when the non-bypass routing approach is used. Here we extend this evaluation for the bypass case to evaluate its impact. In bypass routing, the traffic stays in the optical layer after being converted at the source until its converted electronically back at the destination \cite{Shen2009}. In a bidirectional flow, source and destination nodes use non network coding transponders while intermediate nodes use network coding transponders. Flows between neighbouring nodes use standard non network coding transponders. \\
The MILP model for this scenario uses all the constraints in the previous non-bypass model and decides the optimum number and locations of encoding transponders, and the power consumption of the network for different coded ports power consumption. The objective function takes into consideration changes in the required resources in intermediate nodes as compared to the non-bypass approach. The results in this section are compared with the conventional scenario without network coding as well as the one with network coding in non-bypass approach. We  also carry out a sensitivity analysis to account for the possibility of different power consumption values that the coded port can take in relation to the conventional port and assess the impact of this value on the overall savings.\\
\begin{table}[h]
\small
 \centering 
  \caption{Network Parameters for the bypass case \cite{GreenTouch}} 
 \begin{tabular}{l l} 
 \hline\hline 
 Parameter & Value \\ [0.5ex] 
 \hline 
 Separating distance between EDFAs & 80 km  \\ 
 Wavelengths count in a single fibre (W) & 32  \\
 Capacity of a single wavelength (B) \cite{GreenTouch} & 400 Gbps \\
 Conventional port power consumption (Pp) \cite{CRS1} & 46.7 W  \\
 Coded transponder's power consumption (Px) & 360 kW  \\ 
 Transponder's power consumption (Pt) \cite{transponder} & 332.6 W  \\
Optical Switch's power consumption (PO)&	8.5 W\\
EDFAs power consumption  (Pe)	\cite{EDFA}& 15.3 W\\
 \hline 
 \end{tabular}
 \label{table:parametersBypass}
 \end{table}
The power consumption of the network devices are shown in Table  \ref{table:parametersBypass}. The reason behind using the GreenTouch 2020 power consumption values \cite{GreenTouch} is that the power savings of network coding under the 2010 power consumption values are very small as router ports are the most significant power consumption equipment in the network, which reduces significantly the contribution of savings in transponders due to network coding. However, in 2020, the GreenTouch consortium expects the transponders to consume more power than the router ports which makes network coding savings become more significant. This is because transponders are expected to use more digital signal processing at higher data rates but routers are expected to drop significantly in power consumption due to the use of optical interconnects, and packet size adaptation. To match the projected capacity increase in 2020, a wavelength of capacity 400 Gbps is used.  We also assume the power consumption of the network coding results in network coding ports that consume 10\% and 50\% higher than the conventional ports accounting for the extra required elements. To cover a wide range of this ratio, we show the results for ratios between of 0\% and 100\%. 

\begin{figure}[h]
\centering
\begin{tikzpicture}[scale=1]
\begin{axis} [
xtick=data, grid=both,xmin=0,xmax=22,ymin=0,xlabel={Time of the day},
ylabel={Power (W)},
ylabel near ticks,
legend style={at ={(1,0.35)}}, 
legend cell align=left,  font=\scriptsize
]

\addplot[domain=0:22,samples=12,color=black, mark=o] coordinates
{
(0, 374700	)
(2 ,210100)
(4 ,143500	)
(6 ,122900	)
(8,130500	)
(10,299800	)
(12,401900	)
(14,407400	)
(16,425600	)
(18,456800	)
(20,404500	)
(22,461400)
};
	\addlegendentry{Non bypass}
	
	\addplot[domain=0:22,samples=12,color=blue, mark=square] coordinates
{
(0, 330600)
(2 ,185400)
(4 ,126600)
(6 ,108400)
(8, 115100)
(10,264500)
(12,354600)
(14,359400)
(16,375400)
(18,402900)
(20,356800)
(22,407000)
};
	\addlegendentry{bypass}
	
		\addplot[domain=0:22,samples=12,color=green, mark=x] coordinates
{
(0, 306800	)
(2 ,170900	)
(4 ,117100	)
(6 ,98180	)
(8,	107500)
(10,243200	)
(12,324300	)
(14,330100	)
(16,350100	)
(18,363300	)
(20,323900	)
(22,370700)
};
\addlegendentry{NC bypass (500W)}

\addplot[domain=0:22,samples=12,color=red, mark=x] coordinates
{
(0,278500)
(2 ,156300)
(4 ,103100)
(6 ,87970)
(8,	95690)
(10,223200)
(12,259700)
(14,295400)
(16,309000)
(18,326400)
(20,287800)
(22,333900)
};
\addlegendentry{NC bypass (360W)}

\end{axis} 
\end{tikzpicture}
\caption{Power consumption of the NSFNET with bypass at different time of the day }
\label{fig:NCBypass2}
\end{figure}
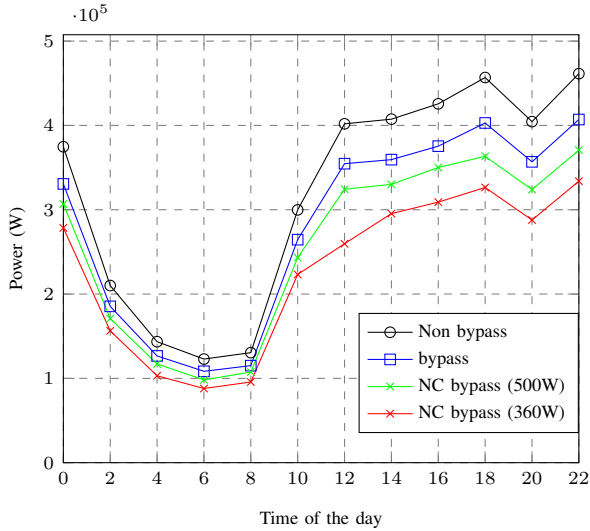

Fig. \ref{fig:NCBypass2} shows that under the bypass approach, savings up to 28\% can be obtained under the NSFNET topology with 10\% increase in coded transponders power consumption and 19\% considering the 50\% increase. These savings are with reference to the non-bypass approach, but when compared to the conventional approach, savings of  18\% and 9\% respectively are obtained for the two investigated transponders power consumption values.\\
\begin{figure}[h]
\centering
\begin{tikzpicture}[scale=1]
\begin{axis} [
xtick=data, grid=both,xmin=330,xmax=660,xlabel={Power per coded transponder (W)},
ylabel={Power (W)},
ylabel near ticks,
legend style={at ={(1.0,0.25)}}, 
legend cell align=left,  font=\scriptsize
]

\addplot[domain=330:660, samples=12,color=blue, mark=square] {374700};
	\addlegendentry{Conventional Non Bypass}
	
	\addplot[domain=330:660, samples=12,color=black, mark=x] {330600};
\addlegendentry{Conventional Bypass}
	
\addplot[domain=0:22,samples=12,color=red, mark=*] coordinates
{
(330,275500)
(360,282200)
(390,284700)
(420,291100)
(450,297800)
(480,302300)
(510,307600)
(540,311900)
(570,316200)
(600,318800)
(630,324700)
(660,330600)
};
\addlegendentry{NC Bypass}

\end{axis} 
\end{tikzpicture}
\caption{Power consumption of the network for different coded transponders' power values for the NSFNET}
\label{fig:bypass}
\end{figure}
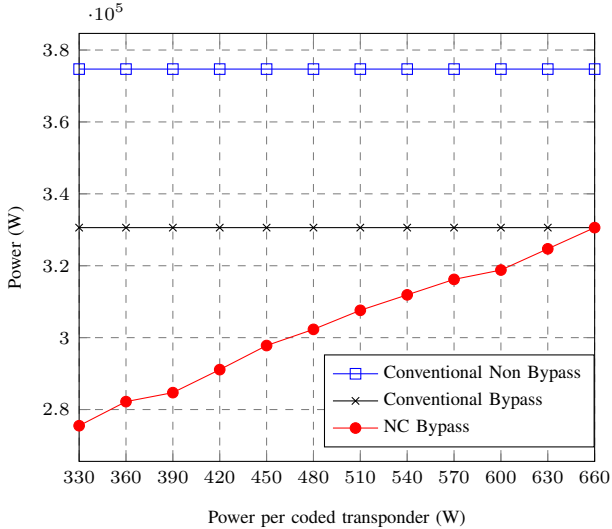
 In Fig. \ref{fig:bypass}, the total power consumption is plotted at different power consumption values for the coded transponder while the conventional transponders power consumption remains fixed to account for the uncertainty in the additional power consumption that is needed to implement the additional functions related to network coding.  This sensitivity analysis is useful in determining the value of the coded transponder power consumption at which the network coding approach seizes to provide savings. The figure shows that the network coding approach with bypass outperforms both the conventional bypass and the conventional non-bypass even if the network coding  transponder consumes 660W at 400 Gbps, and even much higher when compared with the conventional non-bypass case.  \\
In our analysis we considered only the power consumption of the network as the objective function to be minimized. As studied in \cite{wiatr2012power}, network performance can be affected. This opens the door for additional study that takes into account additional metrics (e.g. delay) into the optimization. The MILP model minimizes the network power consumption at different times of the day by optimizing the number of resources needed at each node. This number represents the resources that are switched on at that time. For dimensioning the network the maximum number of each component at each node throughout the day is considered. It is worth noting that we have used multiple traffic matrices at different times of the day to represent the multi-hour traffic  as opposed to relying on the dominant matrices technique in  \cite{pavon2011multi} for network dimensioning. 
\section{analytic bounds and closed forms}
\subsection{Conventional scenario}
	Let $G(\mathcal{N},E)$  be a network composed of a collection of undirected edges, $E$, connecting a set of nodes, $\mathcal{N}$. We define the bidirectional traffic demand as the triple $(s,d,\lambda^{sd})$, where $s \in \mathcal{N}$ and $d \in \mathcal{N}$ are end nodes of the demand and $\lambda^{sd}$ is the data rate requirement of the demand. 
Assuming each demand follows a single path, i.e. no traffic bifurcation, the hop count traversed by a single traffic flow of the bidirectional demand is given by
\begin{equation}\label{analytic1}
    h^{sd}=\sum_{m \in \mathcal{N}}\sum_{n \in \mathcal{N}_m} b^{sd}_{mn},
\end{equation}
where $b_{mn}^{sd}$ is a binary variable, $b_{mn}^{sd}=1$ if the flow between node pair $(s,d)$ is routed through link $(m,n)$, $b_{mn}^{sd}=0$ otherwise.
The average hop count for all flows in the network is given by
\begin{equation}\label{analytic2}
    h=\frac{1}{N(N-1)} \sum_{s \in \mathcal{N}}\sum_{\substack{d \in \mathcal{N} \\s \ne d}} h^{sd}.
\end{equation}
The power consumption of a single traffic flow, $(s,d)$, of the bidirectional demand in a conventional IP over WDM network under the non-bypass approach is given by
\begin{equation}\label{analytic3}
    P_{sd}=(p_p+p_t)\sum_{m \in \mathcal{N}}\sum_{n \in \mathcal{N}_m} \frac{w^{sd}_{mn}}{B}.
\end{equation}
Note that we considered only the most power consuming devices; router ports and transponders for simplicity. Also note that the use of traffic grooming results in efficient wavelength and router resources utilisation and improves the accuracy of (\ref{analytic3}).
Assuming all demands of the network are equal, i.e.  $\lambda^{sd}=\lambda,\forall s,d \in \mathcal{N},s \ne d$,
\begin{equation}\label{analytic4}
\sum_{m \in \mathcal{N}}\sum_{n \in \mathcal{N}_m} w^{sd}_{mn}=\lambda \sum_{m \in \mathcal{N}}\sum_{n \in \mathcal{N}_m} b^{sd}_{m,n}=\lambda h^{sd}.
\end{equation}
Therefore the power consumption of the single flow can be given as
\begin{equation}\label{analytic5}
P_{sd}=\left(\frac{P_t +P_p}{B} \right) \lambda h^{sd}. 
\end{equation}
The total network power consumption can be given as
\begin{equation}\label{analytic6}
P=\left(\frac{P_t +P_p}{B} \right) \lambda \sum_{s \in \mathcal{N}}\sum_{\substack{d \in \mathcal{N} \\ s \ne d}}h^{sd}. 
\end{equation}
Let $P^{sd}_{\lambda} =\left(\frac{P_t +P_p}{B} \right) \lambda^{sd}$  be the single hop power consumption and $P^{sd}_{\lambda} =P_{\lambda}, \forall s,d \in \mathcal{N},s \ne d$; then the power consumption can be given as
\begin{equation}\label{analytic7}
P=P_{\lambda} \sum_{s \in \mathcal{N}}\sum_{\substack{d \in \mathcal{N} \\ s \ne d}}h^{sd}.
\end{equation}
The hop count of a traffic demand in the network can be given as a function of the average hop count $h$ as
\begin{equation}\label{analytic8}
\sum_{s \in \mathcal{N}}\sum_{\substack{d \in \mathcal{N} \\ s \ne d}}h^{sd}=N(N-1)\frac{\sum_{s}\sum_{d}h^{sd}}{N(N-1)}=N(N-1)h. 
\end{equation}
Therefore the power consumption of the IP over WDM network can be given as
\begin{equation}\label{analytic9}
P=P_{\lambda} hN(N-1).
\end{equation}
\subsection{Network coding scenario}
For network coding enabled IP over WDM network, both flows of the bidirectional demand are routed through the same path so network coding is performed at intermediate nodes. In this case the power consumption of the two flows of the bidirectional demand $((s,d)$ and $(d,s))$ is given as
\begin{equation}\label{analytic10}
\tilde{P^{sd}}=2\left( \frac{P_t+P_p}{B}\right)\lambda^{sd}+\left( \frac{P_t+P_x}{B}\right)\lambda^{sd} (h^{sd}-1),
\end{equation}
\centerline{$\forall s,d, \in \mathcal{N}, s < d $}.
The first term in Equation (\ref{analytic10}) gives the power consumption of end nodes where a conventional port is used to send and receive the flows at each end. Note that the XOR gate and storage at the end nodes have small power consumption and are not included here. The second term is attributed to intermediate nodes where coding is implemented. We evaluate equation (\ref{analytic10}) for all values of $s<d$, as opposed to the $s \ne d$ used in the conventional case, because here we take flows in pairs, and hence  $\tilde{P^{sd}}$ accounts for the total power consumed for the flow $(s,d)$ and $(d,s)$. The power consumption can be re arranged as
\begin{equation}\label{analytic11}
\tilde{P^{sd}}=2\lambda^{sd}\frac{P_t+P_p}{B} \left(1+\frac{p_t + p_x }{p_t+p_p} \left( \frac{h^{sd}-1}{2}\right) \right).
\end{equation}
Let $r=\frac{p_t+p_x}{p_t+p_p}$ represent the ratio of the power consumption of the network coding scheme (router ports and transponders) and the conventional scheme, then the power consumption of the bidirectional demand can be written as
\begin{equation}\label{analytic12}
\tilde{P^{sd}}=2 P^{sd}_{\lambda}\left(1+r\frac{h^{sd}-1}{2} \right).
\end{equation}
The total power consumption with network coding is 
\begin{equation}\label{analytic13}
\tilde{P}=2\sum_{s \in \mathcal{N}}\sum_{\substack{d \in \mathcal{N} \\ d <s}}\left( P^{sd}_{\lambda}+P^{sd}_{\lambda} r\frac{h^{sd}-1}{2} \right),
\end{equation}
giving
\begin{equation}\label{analytic14}
\tilde{P}=2\left(\sum_{s \in \mathcal{N}}\sum_{\substack{d \in \mathcal{N} \\ d <s}} P^{sd}_{\lambda} \left( 1-\frac{r}{2} \right) + \frac{r}{2} \sum_{s \in \mathcal{N}}\sum_{\substack{d \in \mathcal{N} \\ d <s}} P^{sd}_{\lambda}h^{sd} \right).
\end{equation}
We start with the equal traffic demands case where all the demands in the network have the same value, i.e. $P^{sd}_{\lambda}=P_{\lambda}, \forall s,d \in \mathcal{N}$. The total power consumption becomes
\begin{equation}\label{analytic15}
\tilde{P}=2 P_{\lambda}\left(\left(1-\frac{r}{2}\right) \left( \frac{N(N-1)}{2}\right)+\frac{rN(N-1)h}{4} \right),
\end{equation}
\begin{equation}\label{analytic16}
\tilde{P}=P_{\lambda} N(N-1) \left(1+\frac{r}{2}(h-1) \right).
\end{equation}
The power savings $\phi$ is then given by
\begin{equation}\label{analytic17}
\phi=1-\frac{\tilde{P}}{P}=1-\left(\frac{P_{\lambda}N(N-1)\left(1+\frac{r(h-1)}{2} \right)}{P_{\lambda} h N(N-1)} \right),
\end{equation}
which gives
\begin{equation}\label{analytic18}
\phi=\left(1-\frac{1+0.5 r (h-1)}{h} \right).
\end{equation}
Using the power consumption values of the conventional and NC ports given in Table \ref{table:parameters}, $r \approx 1.1$, reducing the savings expression to
\begin{equation}\label{analytic19}
\phi=\left(1-\frac{1+0.55 (h-1)}{h} \right).
\end{equation}
\begin{equation}\label{analytic20}
\phi=\left(0.45 \frac{h-1}{h} \right).
\end{equation}
The derived expression is used to calculate the maximum saving gained by implementing network coding in regular topologies (star, ring, line) as the number of nodes grows to infinity. 
\subsubsection{Star topology}
\begin{equation}\label{analytic21}
\phi=\left( 0.45 \frac{\frac{2(N-1)}{N}-1}{\frac{2(N-1)}{N}} \right)=\left(0.45 \frac{N-2}{2(N-1)} \right).
\end{equation}
\begin{equation}\label{analytic22}
\lim_{n \to \infty} \phi=\lim_{n \to \infty}\left( 0.45 \frac{\frac{2(N-1)}{N}-1}{\frac{2(N-1)}{N}} \right)=0.225.
\end{equation}
\subsubsection{Ring topology: Odd number of nodes}
\begin{equation}\label{analytic23}
\phi=\left(0.45 \frac{\frac{N+1}{4}-1}{\frac{N+1}{4}} \right)=\left(0.45 \frac{N-3}{N+1} \right).
\end{equation}
\begin{equation}\label{analytic24}
\lim_{n \to \infty} \phi=\lim_{n \to \infty} \left( 0.45 \frac{N-3}{N+1}\right)=0.45.
\end{equation}
\subsubsection{Ring topology: Even number of nodes}
\begin{equation}\label{analytic25}
\phi=\left(0.45 \frac{\frac{N^2}{4(N-1)}-1}{\frac{N^2}{4(N-1)}}\right)=\left(0.45 (\frac{N-2}{N})^2 \right).
\end{equation}
\begin{equation}\label{analytic26}
\lim_{n \to \infty} \phi=\lim_{n \to \infty} \left( 0.45 (\frac{N-2}{N})^2\right)=0.45.
\end{equation}
\subsubsection{Line topology}
\begin{equation}\label{analytic27}
\phi=\left(0.45 \frac{\frac{N+1}{3}-1}{\frac{N+1}{3}}\right)=\left(0.45 \frac{N-2}{N+1} \right).
\end{equation}
\begin{equation}\label{analytic28}
\lim_{n \to \infty} \phi=\lim_{n \to \infty} \left( 0.45 \frac{N-2}{N+1}\right)=0.45.
\end{equation}
The savings asymptotically reach 45\%  for the ring and line topologies and 22.5\% for the star topology. With network coding ports as power efficient as conventional ports (i.e. $r=1$), the savings increase to 50\% and 25\% respectively.
 \begin{figure}[h]
 \centering
\begin{tikzpicture}[scale=1]
\begin{axis} [
xtick={20,40,60,80,100,120}, grid=both,xmin=0,xmax=120,ymin=0,xlabel={Demand volume (Gbps)},
ylabel={Power (W)},
ylabel near ticks,
legend style={at ={(0.65,1)}}, 
legend cell align=left, font=\small
]
\addplot[color=black, mark=x]
coordinates
{
(0,0)(20,213181)(40,426362)(60,639543)(80,852724)(100,1065905)(120,1279086)
};
\addlegendentry{Conventional (MILP)}

\addplot[color=blue, mark=*]
coordinates
{
(0,0)(20,164304)(40,328608)(60,492912)(80,657216)(100,821520)(120,985824)
};
\addlegendentry{NC (MILP)}
\addplot[color=black, mark=triangle]
coordinates
{
(0,0)(20,174203)(40,348406)(60,522609)(80,696812)(100,871015)(120,1045218)
};
\addlegendentry{NC (Heuristic)}
	\addplot[domain=0:120,color=red] {10594.2655*x };
	\addlegendentry{Conventional (Analytical)}

	\addplot[domain=0:120,color=black,style=dashed] {8023.8135*x };
	\addlegendentry{ NC (Analytical)}
\end{axis} 
\end{tikzpicture}
\caption{Power consumption of the analytical and MILP models for the NSFNET under the zero padding approach }
\label{fig:equalTr}
\end{figure}
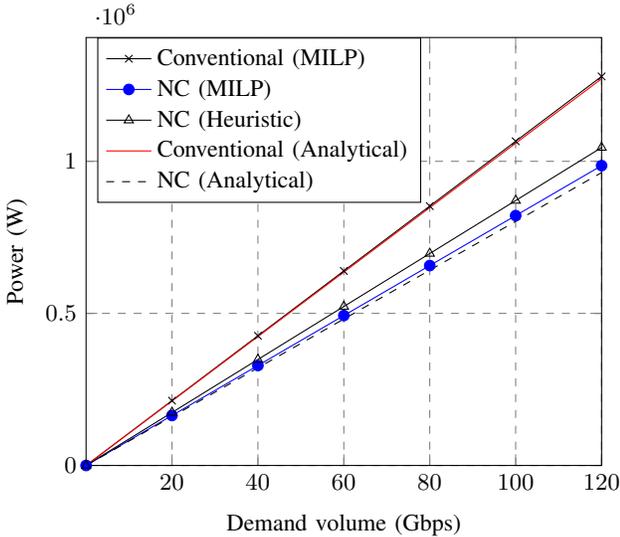
Fig. \ref{fig:equalTr} shows a comparison between the MILP model, the analytical bounds derived and the heuristic for the case of equal  traffic demands between all nodes for the zero padding approach. The small difference between the analytical model and the MILP model is attributed to the simplicity of the analytical formulae which takes into account only the routers and transponders. 
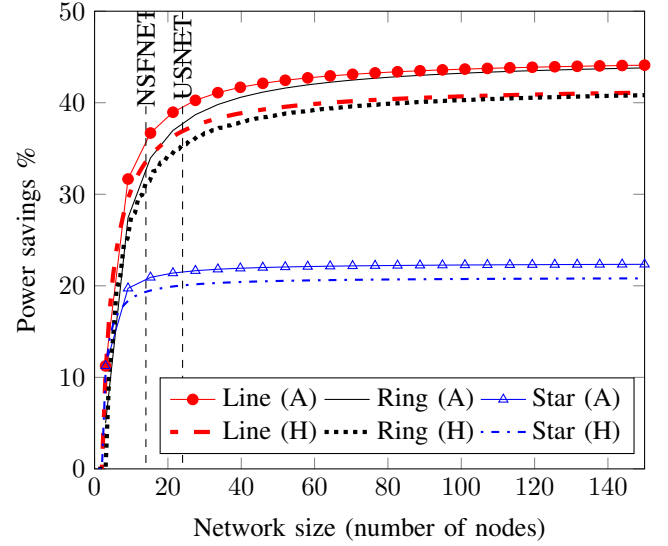
\begin{figure}[h]
\centering
 \begin{tikzpicture}[scale=1]
	\begin{axis}[width=\linewidth,
		xlabel={Network size (number of nodes)},
		ylabel near ticks,
		ylabel={Power savings \%},
		xmin=0,xmax=150,
		ymin=0,ymax=50,
		legend style={at={(0.55,0.2)}, anchor=north, legend columns=3}]
	\addplot[domain=3:150,color=red, mark=*] {45*(x-2)/(x+1)) };
	\addlegendentry{Line (A)}
	\addplot [domain=3:150,color=black]{45*((x-2)/x)^2 };
	\addlegendentry{Ring (A)}
	\addplot [domain=3:150,color=blue, mark=triangle]{45*(x-2)/(2*(x-1)) };
	\addlegendentry{Star (A)}
	\addplot [dashed, ultra thick, color=red, dash pattern=on 3pt off 6pt on 6pt off 6pt,] coordinates{(1	,0)(2	,0)(3	,10.5)(4	,16.7)(5	,21)(6	,24)(7	,26.2)(8	,28)(9	,29.4)(10	,30.50)(11	,31.45)(12	,32.26)(13	,32.95)(14	,33.55)(15	,34.07)(16	,34.53)
(17	,34.94)(18	,35.31)(19	,35.64)(20	,35.94)(21	,36.21)(22	,36.46)(23	,36.69)(24	,36.90)(25	,37.09)(26	,37.27)(27	,37.44)(28	,37.60)(29	,37.74)(30	,37.87)(31	,38.00)(32	,38.12)
(33	,38.23)(34	,38.34)(35	,38.44)(36	,38.53)(37	,38.62)(38	,38.71)(39	,38.79)(40	,38.86)(41	,38.94)(42	,39.01)(43	,39.07)(44	,39.14)(45	,39.20)(46	,39.26)(47	,39.31)(48	,39.37)
(49	,39.42)(50	,39.47)(51	,39.51)(52	,39.56)(53	,39.60)(54	,39.65)(55	,39.69)(56	,39.73)(57	,39.76)(58	,39.80)(59	,39.84)(60	,39.87)(61	,39.90)(62	,39.94)(63	,39.97)(64	,40.00)
(65	,40.03)(66	,40.06)(67	,40.08)(68	,40.11)(69	,40.14)(70	,40.16)(71	,40.19)(72	,40.21)(73	,40.23)(74	,40.26)(75	,40.28)(76	,40.30)(77	,40.32)(78	,40.34)(79	,40.36)(80	,40.38)
(81	,40.40)(82	,40.42)(83	,40.44)(84	,40.45)(85	,40.47)(86	,40.49)(87	,40.50)(88	,40.52)(89	,40.54)(90	,40.55)(91	,40.57)(92	,40.58)(93	,40.60)(94	,40.61)(95	,40.62)(96	,40.64)
(97	,40.65)(98	,40.66)(99	,40.68)(100	,40.69)(101,40.70)(102	,40.71)(103,40.72)(104	,40.74)(105,40.75)(106	,40.76)(107,40.77)(108	,40.78)(109,40.79)(110	,40.80)(111,40.81)(112	,40.82)
(113,40.83)(114	,40.84)(115,40.85)(116	,40.86)(117,40.87)(118	,40.88)(119,40.89)(120	,40.89)(121,40.90)(122	,40.91)(123,40.92)(124	,40.93)(125,40.93)(126	,40.94)(127,40.95)(128	,40.96)
(129,40.97)(130	,40.97)(131,40.98)(132	,40.99)(133,40.99)(134	,41.00)(135,41.01)(136	,41.02)(137,41.02)(138	,41.03)(139,41.03)(140	,41.04)(141,41.05)(142	,41.05)(143,41.06)(144	,41.07)
(145,41.07)(146	,41.08)(147,41.08)(148	,41.09)(149,41.09)(150	,41.10)
	};
	\addlegendentry{Line (H)}
	\addplot [dotted, ultra thick, color=black] coordinates{(1	,0)(2	,0)(3	,0)(4	,10.5)(5	,14)(6	,19.57)(7	,21)(8	,24.5)
(9	,25.2)(10	,27.55)(11	,27.95)(12	,29.70)(13	,29.95)(14	,31.28)(15	,31.45)(16	,32.50)
(17	,32.61)(18	,33.46)(19	,33.55)(20	,34.24)(21	,34.31)(22	,34.89)(23	,34.94)(24	,35.44)
(25	,35.48)(26	,35.91)(27	,35.94)(28	,36.32)(29	,36.34)(30	,36.67)(31	,36.69)(32	,36.98)
(33	,37.00)(34	,37.26)(35	,37.27)(36	,37.51)(37	,37.52)(38	,37.73)(39	,37.74)(40	,37.93)
(41	,37.94)(42	,38.11)(43	,38.12)(44	,38.28)(45	,38.29)(46	,38.43)(47	,38.44)(48	,38.57)
(49	,38.58)(50	,38.70)(51	,38.71)(52	,38.82)(53	,38.83)(54	,38.93)(55	,38.94)(56	,39.04)
(57	,39.04)(58	,39.13)(59	,39.14)(60	,39.22)(61	,39.23)(62	,39.31)(63	,39.31)(64	,39.39)
(65	,39.39)(66	,39.46)(67	,39.47)(68	,39.54)(69	,39.54)(70	,39.60)(71	,39.60)(72	,39.66)
(73	,39.67)(74	,39.72)(75	,39.73)(76	,39.78)(77	,39.78)(78	,39.84)(79	,39.84)(80	,39.89)
(81	,39.89)(82	,39.94)(83	,39.94)(84	,39.98)(85	,39.98)(86	,40.03)(87	,40.03)(88	,40.07)
(89	,40.07)(90	,40.11)(91	,40.11)(92	,40.15)(93	,40.15)(94	,40.19)(95	,40.19)(96	,40.22)
(97	,40.22)(98	,40.26)(99	,40.26)(100	,40.29)(101	,40.29)(102	,40.32)(103	,40.32)(104	,40.35)
(105	,40.35)(106	,40.38)(107	,40.38)(108	,40.41)(109	,40.41)(110	,40.44)(111	,40.44)(112	,40.46)
(113	,40.46)(114	,40.49)(115	,40.49)(116	,40.51)(117	,40.51)(118	,40.54)(119	,40.54)(120	,40.56)
(121	,40.56)(122	,40.58)(123	,40.58)(124	,40.60)(125	,40.60)(126	,40.62)(127	,40.62)(128	,40.64)
(129	,40.64)(130	,40.66)(131	,40.66)(132	,40.68)(133	,40.68)(134	,40.70)(135	,40.70)(136	,40.72)
(137	,40.72)(138	,40.74)(139	,40.74)(140	,40.75)(141	,40.75)(142	,40.77)(143	,40.77)(144	,40.78)
(145	,40.78)(146	,40.80)(147	,40.80)(148	,40.81)(149	,40.82)(150	,40.83)
	};
	\addlegendentry{Ring (H)}
	\addplot [dashed, thick, color=blue, dash pattern=on 1pt off 3pt on 3pt off 3pt] coordinates{
	(1	,0)(2	,0)(3	,10.5)(4	,14)(5	,15.7)(6	,16.7)(7	,17.4)(8	,17.9)
(9	,18.3)(10	,18.63)(11	,18.86)(12	,19.05)(13	,19.21)(14	,19.34)(15	,19.46)(16	,19.56)(17	,19.64)(18	,19.72)(19	,19.79)(20	,19.85)(21	,19.91)(22	,19.96)(23	,20.00)(24	,20.04)
(25	,20.08)(26	,20.12)(27	,20.15)(28	,20.18)(29	,20.21)(30	,20.23)(31	,20.26)(32	,20.28)(33	,20.30)(34	,20.32)(35	,20.34)(36	,20.36)(37	,20.37)(38	,20.39)(39	,20.40)(40	,20.42)
(41	,20.43)(42	,20.44)(43	,20.46)(44	,20.47)(45	,20.48)(46	,20.49)(47	,20.50)(48	,20.51)(49	,20.52)(50	,20.53)(51	,20.54)(52	,20.54)(53	,20.55)(54	,20.56)(55	,20.57)(56	,20.57)
(57	,20.58)(58	,20.59)(59	,20.59)(60	,20.60)(61	,20.61)(62	,20.61)(63	,20.62)(64	,20.62)(65	,20.63)(66	,20.63)(67	,20.64)(68	,20.64)(69	,20.65)(70	,20.65)(71	,20.66)(72	,20.66)
(73	,20.66)(74	,20.67)(75	,20.67)(76	,20.68)(77	,20.68)(78	,20.68)(79	,20.69)(80	,20.69)(81	,20.69)(82	,20.70)(83	,20.70)(84	,20.70)(85	,20.70)(86	,20.71)(87	,20.71)(88	,20.71)
(89	,20.72)(90	,20.72)(91	,20.72)(92	,20.72)(93	,20.73)(94	,20.73)(95	,20.73)(96	,20.73)(97	,20.74)(98	,20.74)(99	,20.74)(100,20.74)(101,20.74)(102,20.75)(103,20.75)(104,20.75)
(105,20.75)(106,20.75)(107,20.76)(108,20.76)(109,20.76)(110,20.76)(111,20.76)(112,20.77)(113,20.77)(114,20.77)(115,20.77)(116,20.77)(117,20.77)(118,20.78)(119,20.78)(120,20.78)
(121,20.78)(122,20.78)(123,20.78)(124,20.78)(125,20.79)(126,20.79)(127,20.79)(128,20.79)(129,20.79)(130,20.79)(131,20.79)(132,20.79)(133,20.80)(134,20.80)(135,20.80)(136,20.80)
(137,20.80)(138,20.80)(139,20.80)(140,20.80)(141,20.80)(142,20.81)(143,20.81)(144,20.81)(145,20.81)(146,20.81)(147,20.81)(148,20.81)(149,20.81)(150,20.81)
	};
	\addlegendentry{Star (H)}
\draw[dashed] ({axis cs:14,0}|-{rel axis cs:0,1}) -- ({axis cs:14,0}|-{rel axis cs:0,0});
\node at (axis cs:14,45)[rotate=90]{NSFNET};
\node at (axis cs:24,45)[rotate=90]{USNET};
\draw[dashed] ({axis cs:24,0}|-{rel axis cs:0,1}) -- ({axis cs:24,0}|-{rel axis cs:0,0});
	\end{axis}
\end{tikzpicture}
\caption{Effect of network coding on large network sizes (A:analytical and H:Heuristic)}
\label{fig:LargeNetworks}
\end{figure}

Fig. \ref{fig:LargeNetworks} shows the power savings as calculated by the analytical formulae for the line, ring and star topologies up to a network size of 150 nodes and compares these savings to the savings of the minimum hop routing heuristic for the network coding enabled networks for the zero padding approach. For each network size, the topologies are built with link distances equal to the average NSFNET link distance. The figure also provides results for the NSFNET and USNET to show an example of where common core networks are located. The analytically calculated power savings confirm those of the heuristic. The savings saturate just below 45\% and 22.5\% as the size increases. The figure shows that the ring topology converges to the maximum value faster than the line topology.  Note that the power consumption of the analytical results does not take into account the power consumption of EDFAs, multiplexers and optical switches; therefore the power savings of the heuristic are lower than the analytically calculated power savings.\par
For the general case when the bidirectional traffic demand volume is randomly distributed we prove first that the previous case of equal average traffic demands has better power efficiency compared to the random demands case. Considering the case of random traffic demands, equation (\ref{analytic13}) becomes
\begin{equation}\label{analytic29}
\tilde{P}=\sum_{s \in \mathcal{N}}\sum_{\substack{d \in \mathcal{N} \\ d <s}}\bigg[ 2 \max(P^{sd}_{\lambda},P^{ds}_{\lambda})+\max(P^{sd}_{\lambda},P^{ds}_{\lambda}) r(h^{sd}-1) \bigg].
\end{equation}
The first part of equation (\ref{analytic29}) accounts for the power consumption at end nodes which uses conventional ports, while the second part accounts for the power consumption at intermediate nodes which uses NC ports. We treat demands in pairs and hence the use of $(d < s)$ under the summation and the reason for the multiplication by $2$.  \par
The value of $\max(P^{sd}_{\lambda},P^{ds}_{\lambda})$ can be written as
\begin{equation}\label{analytic30}
\max(P^{sd}_{\lambda},P^{ds}_{\lambda})=\left(\frac{p_t+p_x}{B} \right) \max(\lambda^{sd},\lambda^{ds}).
\end{equation}
\begin{multline}\label{analytic31}
\max(\lambda^{sd},\lambda^{ds})=\max(\lambda+\Delta^{sd},\lambda+\Delta^{ds})\\=\lambda+\max(\Delta^{sd},\Delta^{ds}),
\end{multline}
where $\lambda$ is the average traffic of all the network.
The total power of the NC case can be written as
\begin{equation}\label{analytic32}
\tilde{P}=2\left(\frac{p_t+p_p}{B} \right) \sum_{\substack{s,d \in \mathcal{N}\\ d <s}}\bigg[ \lambda+\max(\Delta^{sd},\Delta^{ds})\bigg] \bigg[1+r\frac{h^{sd}-1}{2} \bigg]
\end{equation}
which can be written as
\begin{multline}\label{analytic33}
\tilde{P}=P_{\lambda}N(N-1) \bigg[ 1+r\frac{h-1}{2} \bigg]+\\2\left(\frac{p_t+p_p}{B} \right)\sum_{s \in \mathcal{N}}\sum_{\substack{d \in \mathcal{N}\\ d <s}}\max(\Delta^{sd},\Delta^{ds}) \bigg[1+r\frac{h^{sd}-1}{2} \bigg].
\end{multline}
The first two components of equation (\ref{analytic33}) are the power consumption of the network coded case, when the traffic demands are all equal to the average. We call it $\tilde{P^{(1)}}$ and we call the second part $\tilde{P^{(2)}}$. Therefore
\begin{equation}\label{analytic34}
\tilde{P^{(1)}}=P_{\lambda}N(N-1) \left( 1+r \frac{h-1}{2}\right).
\end{equation}
\begin{equation}\label{analytic_35}
\tilde{P^{(2)}}=2 \left(\frac{p_t+p_p}{B} \right)\sum_{\substack{s,d \in \mathcal{N}\\ d <s}}\left(\max(\Delta^{sd},\Delta^{ds}) \bigg[1+ r\frac{h^{sd}-1}{2} \bigg] \right).
\end{equation}
The value $\tilde{P^{(1)}}$ represents the total power consumption of the network coding case when all demands are equal, the same as given by equation (\ref{analytic16}). \par
By dividing the set of all demands into subsets, each identified by the number of hops they take, i.e. $H_k$ is the set of demands with minimum hop paths of $k$ hops, then equation (\ref{analytic_35}) can be written as
\begin{multline}\label{analytic36}
\tilde{P^{(2)}}= \frac{p_t+p_p}{B} \bigg[ r \sum_{\substack{(s,d) \in H_2\\ d <s}}\max(\Delta^{sd},\Delta^{ds})+\\
2r \sum_{\substack{(s,d) \in H_3\\ d <s}}\max(\Delta^{sd},\Delta^{ds})+...+\\
(k-1)r \sum_{\substack{(s,d) \in H_k\\ d < s}} \max(\Delta^{sd},\Delta^{ds}) \bigg]+\\2\frac{p_t+p_p}{B}\sum_{s \in N}\sum_{\substack{d \in \mathcal{N}\\ d <s}} \max(\Delta^{sd},\Delta^{ds}).
\end{multline}
Let $g_z(\lambda^{sd})=\max(\Delta^{sd},\Delta^{ds})$, then 
\begin{multline}\label{analytic37}
\tilde{P^{(2)}}= \left(\frac{p_t+p_p}{B} \right)  \bigg[\sum_{k}\sum_{\substack{(s,d) \in H_k \\ d < s}}(k-1)r g_z(\lambda^{sd}) \\+ 2 \sum_{s \in \mathcal{N}}\sum_{\substack{d \in \mathcal{N}\\ d < s}} g_z(\lambda^{sd})\bigg].
\end{multline}
The value of $\tilde{P^{(2)}}$ depends on the given topology (reflected in $H_k$) and the given traffic volume distribution $g_z(\lambda^{sd})$. This produces three lower bounds , one by setting all hop counts to the minimum, another by setting the traffic to a value that minimises the total power, and a third by setting the hop count and the traffic components to their minimum values. The same applies for the three upper bounds. \par
 The bounds for the total power $\tilde{P}$ are as follows:
For a given topology, the minimum value is when $g_z(\lambda^{sd})=\max(\Delta^{sd},\Delta^{ds})=0$ when $\Delta^{sd}=\Delta^{ds}=0$. These values are attained when demands are equal. This gives the following expression of $\tilde{P}$ as
\begin{equation}\label{analytic38}
\tilde{P} \ge P_{\lambda} N (N-1) \left(1+r\frac{h-1}{2} \right), 
\end{equation}
reducing the case to the previous case of the equal average demands. \par
For the optimal topology and generic traffic demands, we attain the following minimum value, when all demands have a single hop route $(i.e. h=1)$ when the network is connected in full mesh:
\begin{equation}\label{analytic39}
\tilde{P} \ge P_{\lambda} N (N-1).
\end{equation}
This means, the higher the variation, the higher the power consumption. Also, when given a set of traffic demands with a given variation, the lowest power consumption will be when bidirectional demands with the highest variance happen to be allocated the route with the minimum hop count. \par
Likewise, We can also find the maximum value of $\tilde{P^{(2)}}$ by considering the topology and traffic dimensions. Considering the traffic dimension, starting from equation (\ref{analytic29}) and  by using the fact that $max(P^{sd}_{\lambda},P^{ds}_{\lambda}) \le \frac{p_t+p_p}{B} \lambda_{max}$ where $\lambda_{max}$ is the upper limit to the traffic value, assuming uniform traffic distribution 

\begin{equation}\label{analytic29-2}
\tilde{P} \le 2 \frac{p_t+p_p}{B} \lambda_{max} \sum_{s \in \mathcal{N}}\sum_{\substack{d \in \mathcal{N} \\ d <s}}\bigg[1+ r\frac{h^{sd}-1}{2} \bigg],
\end{equation}

this gives
\begin{equation}\label{analytic29-3}
\tilde{P} \le  \frac{p_t+p_p}{B} \lambda_{max} N(N-1) \bigg[1+ r\frac{h-1}{2} \bigg].
\end{equation}






An upper bound considering the maximum hop count and the exact traffic is given by setting the hop count for each demand to the maximum in the network, i.e.  $h^{s,d}=h_{max}$ in equation (\ref{analytic_35}), which gives 
\begin{equation}\label{analytic35}
\tilde{P^{(2)}} \le 2 \left(\frac{p_t+p_p}{B} \right)  \bigg[1+ r\frac{h_{max}-1}{2} \bigg]\sum_{s \in \mathcal{N}}\sum_{\substack{d \in \mathcal{N}\\ d <s}}g_z(\lambda^{sd}).
\end{equation}
Therefore this bound for the total power consumption becomes
\begin{multline}\label{analytic42}
\tilde{P} \le P_{\lambda}N(N-1) \left(1+r\frac{h-1}{2} \right)+\\2 \left(\frac{p_t+p_p}{B} \right)  \bigg[1+ r\frac{h_{max}-1}{2} \bigg]\sum_{s \in \mathcal{N}}\sum_{\substack{d \in \mathcal{N}\\ d <s}}g_z(\lambda^{sd}).
\end{multline}
Considering both the maximum traffic and hop count, we have the following upper bound:
\begin{equation}\label{analytic29-3}
\tilde{P} \le  \frac{p_t+p_p}{B} \lambda_{max} N(N-1) \bigg[1+ r\frac{h_{max}-1}{2} \bigg].
\end{equation}
The upper bound given by considering the maximum hop count is tighter than the one considering the maximum possible traffic demand, due to the lower variance the hop count has compared to the traffic demands variance. For the partitioning approach, we develop closed form expression the same way we did with the zero padding approach. The number of NC ports $X$ in the network for the partitioning case is given by
\begin{equation}\label{analytic44}
X=\frac{1}{B}\sum_{s \in \mathcal{N}}\sum_{\substack{d \in \mathcal{N} \\ d < s}}\min(\lambda^{sd},\lambda^{ds})(h^{sd}-1).
\end{equation}
The number of conventional ports will be covering the traffic at source and destination nodes and the remaining traffic of the partitioning process at intermediate nodes. 
\begin{equation}\label{analytic45}
Y=\frac{1}{B} \left(\sum_{s \in \mathcal{N}}\sum_{\substack{d \in \mathcal{N} \\ d \ne s}}\lambda^{sd}+ \sum_{s \in N}\sum_{\substack{d \in \mathcal{N} \\ d < s}} (h^{sd}-1) | \lambda^{sd}-\lambda^{ds}|)\right).
\end{equation}
So the total power consumption will be
\begin{multline}\label{analytic46}
P_t =\frac{p_x+p_t}{B}\sum_{s \in \mathcal{N}}\sum_{\substack{d \in \mathcal{N}\\ d<s}}\min(\lambda^{sd},\lambda^{ds})(h^{sd}-1)+\\\frac{p_p+p_t}{B}\sum_{s \in \mathcal{N}}\sum_{\substack{d \in \mathcal{N} \\ d \ne s}} \lambda^{sd}+\\\frac{p_p+p_t}{B}\sum_{s \in \mathcal{N}}\sum_{\substack{d \in \mathcal{N}\\ s<d}} (h^{sd}-1)|\lambda^{sd}-\lambda^{ds}|.
\end{multline}
Terms can be combined together to give
\begin{multline}\label{analytic47}
P_t =\frac{p_p+p_t}{B}\sum_{s \in \mathcal{N}}\sum_{\substack{d \in \mathcal{N}\\ d \ne s}}\lambda^{sd}+\\
\sum_{s \in \mathcal{N}}\sum_{\substack{d \in \mathcal{N} \\ d < s}} (h^{sd}-1)\bigg(\frac{p_x + p_t}{B} \min(\lambda^{sd},\lambda^{ds}) +\\ \frac{p_p+p_t}{B}|\lambda^{sd}-\lambda^{ds}| \bigg).
\end{multline}
By arranging terms we get
\begin{multline}\label{analytic48}
P_t =\frac{p_p+p_t}{B}N(N-1)\lambda +\\
\frac{p_p+p_t}{B}\sum_{s \in \mathcal{N}}\sum_{\substack{d \in \mathcal{N} \\ d < s}} (h^{sd}-1)\bigg( r \min(\lambda^{sd},\lambda^{ds}) + \\
|\lambda^{sd}-\lambda^{ds}| \bigg).
\end{multline}
\begin{multline}\label{analytic49}
P_t =\frac{p_p+p_t}{B}N(N-1)\lambda +\\
\frac{p_p+p_t}{B}\sum_{s \in \mathcal{N}}\sum_{\substack{d \in \mathcal{N} \\ d < s}} (h^{sd}-1)\bigg[\Delta+r \min(\lambda^{sd},\lambda^{ds}) \bigg].
\end{multline}
where $\Delta=|\lambda^{sd}-\lambda^{ds}|=\max(\lambda^{sd},\lambda^{ds})-\min(\lambda^{sd},\lambda^{ds})$.  When this is substituted in equation (\ref{analytic49}) it gives
\begin{multline}\label{analytic50}
P_t =\frac{p_p+p_t}{B}N(N-1)\lambda +
\frac{p_p+p_t}{B}\sum_{s \in \mathcal{N}}\sum_{\substack{d \in \mathcal{N} \\ d < s}} ( h^{sd}-1) \\\bigg[\max(\lambda^{sd},\lambda^{ds})-\min(\lambda^{sd},\lambda^{ds}) +r\min(\lambda^{sd},\lambda^{ds}) \bigg].
\end{multline}
Grouping similar terms gives
\begin{multline}\label{analytic51}
P_t =\frac{p_p+p_t}{B}N(N-1)\lambda +\\
\frac{p_p+p_t}{B}\sum_{s \in \mathcal{N}}\sum_{\substack{d \in \mathcal{N} \\ d < s}} (h^{sd}-1)\bigg[ \max(\lambda^{sd},\lambda^{ds})+\\(r-1) \min(\lambda^{sd},\lambda^{ds}) \bigg].
\end{multline}
Let the function $g_p(\lambda^{sd})$ be defined as
\begin{equation}\label{analytic52}
g_p(\lambda^{sd})=\max(\lambda^{sd},\lambda^{ds})+(r-1) \min(\lambda^{sd},\lambda^{ds}).
\end{equation}

The function $g_p(\lambda^{sd})$ represents the maximum traffic imbalance in a network where the network coding ports and the conventional ports consume the same power. \par 
The total power becomes
\begin{equation}\label{analytic56}
P_t =\frac{p_p+p_t}{B}N(N-1)\lambda +
\frac{p_p+p_t}{B}\sum_{s \in \mathcal{N}}\sum_{\substack{d \in \mathcal{N} \\ s < d}} (h^{sd}-1)g_p(\lambda^{sd}).
\end{equation}
The lower bound considering the traffic dimension is found from minimising equation (\ref{analytic52})
\begin{equation}\label{analytic56-2}
g_{p_{min}}(\lambda^{sd})= \bar{\lambda}^{sd}+(r-1)\bar{\lambda}^{sd}=r\bar{\lambda}^{sd},
\end{equation}
where $\bar{\lambda}^{sd}$ is the traffic volume between $(s,d)$ when the maximum value equals the minimum value and the average. The total power then becomes
\begin{equation}\label{analytic56-3}
P_t =\frac{p_p+p_t}{B}N(N-1)\lambda +
\frac{p_p+p_t}{B}\sum_{s \in \mathcal{N}}\sum_{\substack{d \in \mathcal{N} \\ s < d}} (h^{sd}-1)\bar{\lambda}^{sd} r.
\end{equation}
By using Chebyshev's inequality shown in (\ref{chebyshev}) where a lower bound on the average of the inner product of two vectors of size $n$ is 
\begin{equation}\label{chebyshev}
\frac{1}{n} \sum_{k=1}^{n} a_k b_k \ge \left( \frac{1}{n}\sum_{k=1}^{n} a_k \right)\left( \frac{1}{n}\sum_{k=1}^{n} b_k \right),
\end{equation}
then equation \ref{analytic56-3} becomes
\begin{multline}\label{analytic56-4}
P_t  \ge \frac{p_p+p_t}{B}N(N-1)\lambda \\+
\frac{p_p+p_t}{B} \sum_{k=1}^{\frac{N(N-1)}{2}}\frac{1}{\frac{N(N-1)}{2}} (h^{k}-1)\sum_{k=1}^{\frac{N(N-1)}{2}}r \bar{\lambda}^{k}.
\end{multline}

By further reducing the second term of the inequality (\ref{analytic56-4}) we get
\begin{multline}\label{analytic56-5}
P_t  \ge \frac{p_p+p_t}{B}N(N-1)\lambda \\+
\frac{p_p+p_t}{B} \bigg[ \frac{1}{\frac{N(N-1)}{2}} \left( h \frac{N(N-1)}{2} - \frac{N(N-1)}{2}\right) r \lambda \frac{N(N-1)}{2} \bigg].
\end{multline}
This is reduced to
\begin{equation}\label{analytic57}
P_t  \ge \frac{p_p+p_t}{B}N(N-1)\lambda \bigg[ 1+ (h-1) \frac{r}{2}
\bigg].
\end{equation}

When we consider the topology dimension
\begin{equation}\label{analytic59}
P_t \ge P_{\lambda} N(N-1)+\frac{p_p+p_t}{B}(h_{min}-1)\sum_{s \in \mathcal{N}}\sum_{\substack{d \in \mathcal{N}\\s < d}}g_p(\lambda^{sd}),
\end{equation}
since $h_{min}=1$
\begin{equation}\label{analytic60}
P_t \ge P_{\lambda} N(N-1).
\end{equation}
For the upper bound, considering the topology dimension
\begin{equation}\label{analytic61}
P_t \le P_{\lambda} N(N-1)+\frac{p_p+p_t}{2B}(h_{max}-1)\sum_{s \in \mathcal{N}}\sum_{\substack{d \in \mathcal{N}\\s < d}}g_p(\lambda^{sd}),
\end{equation}
and considering the traffic dimension, using the fact that the maximum power consumption of the network under network coding occurs when network coding is not used. Under the partitioning approach, network coding is not used when the bidirectional traffic is fully asymmetric, i.e. in one direction $\lambda^{sd}=\lambda_{max}$ and in the other $\lambda^{sd}=0$. Then equation (\ref{analytic56}) becomes:
\begin{equation}\label{analytic62}
P_t \le P_{\lambda} N(N-1)+\frac{p_p+p_t}{B} \lambda_{max} \sum_{s \in \mathcal{N}}\sum_{\substack{d \in \mathcal{N}\\s < d}} (h^{sd}-1),
\end{equation}
which gives:
\begin{equation}\label{analytic62}
P_t \le N(N-1)\frac{p_p+p_t}{B} \bigg[ \lambda+ \lambda_{max} \frac{h-1}{2}\bigg].
\end{equation}
When we consider both the largest hop count and the maximum traffic in the network we get the following upper bound in (\ref{analytic64}). This upper bound is important as the previous two special upper bounds (considering the topology and considering the traffic separately) can have a varying performance. The bound with the max traffic volume is generally higher than the one with the maximum hop count, but in the case of a network with a flat traffic and a very large hop count the opposite occurs. 
\begin{equation}\label{analytic63}
P_t \le P_{\lambda} N(N-1)+\frac{p_p+p_t}{B}(h_{max}-1). \lambda_{max} \frac{N(N-1)}{2}
\end{equation}
\begin{equation}\label{analytic64}
P_t \le  N(N-1)\frac{p_p+p_t}{B} \bigg[\lambda+ \lambda_{max} \frac{h_{max}-1}{2} \bigg].
\end{equation}
We notice that the power consumption given by network coding with traffic partitioning approach under equal traffic demands between all node pairs given by equation (\ref{analytic59}) is the same as the one given by the zero padding case with equal traffic demands given by equation (\ref{analytic59}). Also the lower bounds are the same in both cases, when optimising the topology, giving the full mesh which produces no contribution from network coding, or the minimal traffic case when both bidirectional flows are equal to the average. \par
As mentioned above, the upper bound given by considering the maximum hop count is tighter than the one considering the maximum possible traffic demand. \\
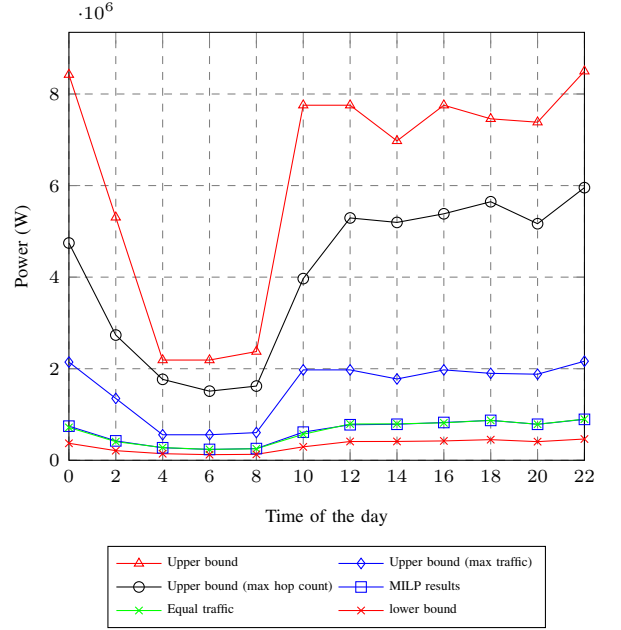
\begin{figure}[h]
\centering
\begin{tikzpicture}[scale=1]
\begin{axis} [
xtick=data, grid=both,xmin=0,xmax=22,ymin=0,xlabel={Time of the day},
ylabel={Power (W)},
ylabel near ticks,
legend style={
			at={(0.5,-0.2)},
			anchor=north,
			legend columns=2, font =\tiny},
legend cell align=left,  font=\scriptsize
]

\addplot[domain=0:22,samples=12,color=red, mark=triangle] coordinates
{
(0, 8422685)
(2 ,5305920)
(4 ,2189156)
(6 ,2189156)
(8,	2374677)
(10,7754807)
(12,7754807)
(14,6975615)
(16,7754807)
(18,7457972)
(20,7383763)
(22,8496893)
};
	\addlegendentry{Upper bound}

\addplot[domain=0:22,samples=12,color=blue, mark=diamond] coordinates
{
(0, 2144459 )
(2 ,1350915 )
(4 ,557370  )
(6 ,557370  )
(8,	604605  )
(10,1974414 )
(12,1974414 )
(14,1776028 )
(16,1974414 )
(18,1898839 )
(20,1879945 )
(22,2163353 )
};
	\addlegendentry{Upper bound (max traffic)}
	
\addplot[domain=0:22,samples=12,color=black, mark=o] coordinates
{

(0, 4745887)
(2 ,2735282)
(4 ,1767607)
(6 ,1509961)
(8,	1620836)
(10,3967977)
(12,5292833)
(14,5195061)
(16,5383665)
(18,5646792)
(20,5166765)
(22,5954453)
};
	\addlegendentry{Upper bound (max hop count)}
	
	\addplot[domain=0:22,samples=12,color=blue, mark=square] coordinates
{
(0,    745533)
(2 ,      423793)
(4 ,      273449)
(6 ,      237221)
(8,       255551)
(10,      614628)
(12,      776080)
(14,      786713)
(16,      825626)
(18,      871232)
(20,      786506)
(22,      894010)
};
	\addlegendentry{MILP results}
	
		\addplot[domain=0:22,samples=12,color=green, mark=x] coordinates
{
(0, 717824)
(2 ,407086)
(4 ,278540)
(6 ,237812)
(8,	251645)
(10,568649)
(12,793014)
(14,795954)
(16,818608)
(18,872156)
(20,787230)
(22,899292)
};
\addlegendentry{Equal traffic}
	
\addplot[domain=0:22,samples=12,color=red, mark=x] coordinates
{
(0,       370968)
(2 ,      210380)
(4 ,      143948)
(6 ,      122900)
(8,	      130049)
(10,      293875)
(12,      409826)
(14,      411346)
(16,      423053)
(18,      450727)
(20,      406837)
(22,      464750)
};
\addlegendentry{lower bound}

\end{axis} 
\end{tikzpicture}
\caption{MILP results, the upper and lower bounds of the zero padding network coding approach }
\label{fig:bounds1}
\end{figure}
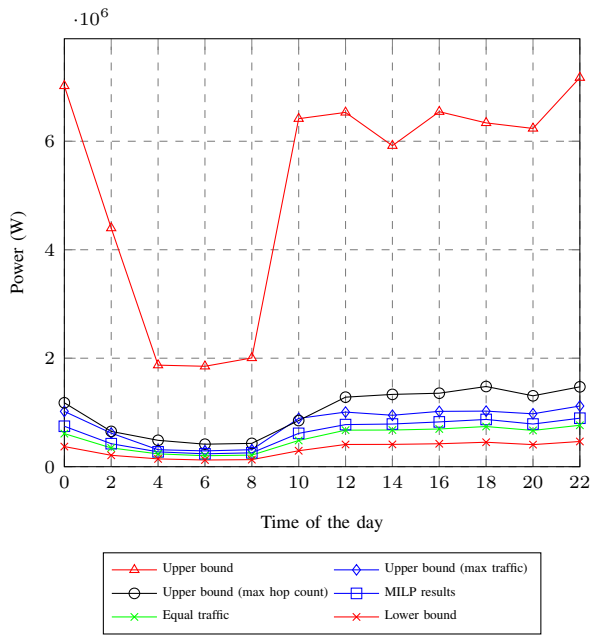
\begin{figure}[h]
\centering
\begin{tikzpicture}[scale=1]
\begin{axis} [
xtick=data, grid=both,xmin=0,xmax=22,ymin=0,xlabel={Time of the day},
ylabel={Power (W)},
ylabel near ticks,
legend style={
			at={(0.5,-0.2)},
			anchor=north,
			legend columns=2, font =\tiny},
legend cell align=left,  font=\scriptsize
]

\addplot[domain=0:22,samples=12,color=red, mark=triangle] coordinates
{
(0, 7020456)
(2 ,4399265)
(4 ,1872229)
(6 ,1851181)
(8,	2004795)
(10,6416091)
(12,6532042)
(14,5918411)
(16,6545269)
(18,6338599)
(20,6236124)
(22,7172824)
};
	\addlegendentry{Upper bound}

\addplot[domain=0:22,samples=12,color=blue, mark=diamond] coordinates
{
(0, 1019293 )
(2 ,618797  )
(4 ,312456  )
(6 ,291408  )
(8,	312837  )
(10,890791  )
(12,1006742 )
(14,948284  )
(16,1019969 )
(18,1024794 )
(20,975193  )
(22,1118788 )
};
	\addlegendentry{Upper bound (max traffic)}

\addplot[domain=0:22,samples=12,color=black, mark=o] coordinates
{
(0, 1177971)
(2 ,650740 )
(4 ,486289 )
(6 ,414863 )
(8,	429899 )
(10,850923 )
(12,1281692)
(14,1332784)
(16,1354631)
(18,1478392)
(20,1306869)
(22,1472297)
};
	\addlegendentry{Upper bound (max hop count)}
	
	\addplot[domain=0:22,samples=12,color=blue, mark=square] coordinates
{
(0,   745533  )
(2 ,  423793  )
(4 ,  273449  )
(6 ,  237221  )
(8,   255551  )
(10,  614628  )
(12,  776080  )
(14,  786713  )
(16,  825626  )
(18,  871232  )
(20,  786506  )
(22,  894010  )
};
	\addlegendentry{MILP results}
	
		\addplot[domain=0:22,samples=12,color=green, mark=x] coordinates
{
(0, 609687)
(2 ,345760)
(4 ,236579)
(6 ,201987)
(8,	213736)
(10,482984)
(12,673550)
(14,676047)
(16,695288)
(18,740769)
(20,668637)
(22,763817)
};
\addlegendentry{Equal traffic}
	
\addplot[domain=0:22,samples=12,color=red, mark=x] coordinates
{
(0, 370968)
(2 ,210380)
(4 ,143948)
(6 ,122900)
(8,	130049)
(10,293875)
(12,409826)
(14,411346)
(16,423053)
(18,450727)
(20,406837)
(22,464750)
};
\addlegendentry{Lower bound}

\end{axis} 
\end{tikzpicture}
\caption{MILP results, the upper and lower bounds of the partitioning network coding approach}
\label{fig:bounds2}
\end{figure}

Figures (\ref{fig:bounds1}) and (\ref{fig:bounds2}) show the upper and lower bounds along side the MILP results of the power consumption of network coding under the zero padding and partitioning approaches respectively. The MILP results lie between these bounds. The power consumption under equal traffic demands which are equal to the average traffic demand is shown to be an approximate representation of the case of random traffic demands. This is due to the fact that multiple demands end up having comparable volumes in most cases, and also because the aggregate bidirectional flows passing through a node, from multiple demands reduce traffic variation and thus the traffic approaches the equal demand case,   making the equal traffic formulas of significant value. However, if the MILP was used in a case where one data centre dominates for example, the MILP and the analytical case with equal average traffic volumes may not agree. 
\section{Conclusions}
In this work, we introduced a technique that can be used to improven energy efficiencyt technique forin IP over WDM networks using network coding. The idea presented here proposes a departure from the conventional router ports, and offers a new architecture that encodes bidirectional flows using a simple xor operation. In order to evaluate the potential power savings, we formulated a MILP model, with the objective of minimising the operational power and developed a minimum hop count routing heuristic for our network coding approach. The results suggest that network coding improves power efficiency, as daily average savings of 27\% and 33\% are obtained when the networks evaluated were NSFNET and USNET respectively. We investigated the impact of network topology on the savings by first replacing the NSFNET by line, ring, star and full mesh topologies. The highest savings are obtained in the line topology (33\%) as a result of the high average hop count of this topology. Network coding offers no energy saving in the full mesh topology (which has an average hop count of 1) due to the unavailability of intermediate nodes that can perform network coding. The minimal contribution of network coding for the star topology is due to its low average hop count (1.85). It is also shown that energy efficient routing protocols in the conventional approach are portable to the network coded approach as the minimum hop routing heuristic is used to route traffic flows in network coding enabled IP over WDM networks where network coding is performed in all intermediate nodes traversed by bidirectional traffic flows. The power savings gained by the heuristic approach match those obtained by the MILP model. We also presented a sensitivity analysis showing the impact on savings as a result of varying the amount of power the network coded ports consume to account for the uncertainty in our estimation. The highest energy efficiency is obtained when the network coded port consumes the same power as the conventional port while losses can be encountered if the NC ports power consumption exceeds twice the power of the conventional port under typical parameters. We analysed two approaches for coding imbalanced bidirectional traffic. The first pads the smaller flow with zeros and the second partitions the larger flow into two components one of them the size of the smaller packet and hence gets coded; while the other component is routed in a conventional manner. We showed that the packet partitioning approach is superior to the zero padding approach especially under asymmetric traffic. We also studied the implementation of network coding when the network employs bypass routing. In this case network coding is performed in the optical layer. Rather than saving resources at the IP layer, resources are saved in the transponders of the intermediate nodes. We have shown that implementing network coding in the optical layer (bypass case) offers less energy savings when compared to the IP layer implementation (nonbypass case) as the IP routers are the highest power consuming devices in the network. In contrast, when considering the expected improvement in router ports power consumption efficiency as suggested by Greentouch and the rise in power consumption of transponders, the savings of the bypass approach become more significant.
Analytic closed form expressions and
bounds are derived and the MILP results are verified. The
formulas are developed for the zero padding and partitioning
network coding approaches for the cases of equal traffic
volumes as well as the uniformly distributed traffic demands.
These formulas are used to study regular networks as their
size grows.
\section*{Acknowledgments}
This authors would like to acknowledge support from the Engineering and Physical Sciences Research Council (EPSRC), INTERNET (EP/H040536/1), STAR (EP/K016873/1) and TOWS (EP/S016570/1) projects. All data are provided in full in the results section of this paper.

\ifCLASSOPTIONcaptionsoff
  \newpage
\fi



\bibliographystyle{IEEEtran}
\bibliography{main.bib}
%



%

\begin{IEEEbiographynophoto}{Mohamed Musa}
received the BSc degree (first-class Honours) in Electrical and Electronic Engineering from the University of Khartoum, Sudan, in 2009, the MSc degree (with distinction) in Broadband Wireless and Optical Communication from University of Leeds, UK, in 2011. He received the PhD from University of Leeds in 2016 in energy efficient network coding in optical networks. He is currently a postdoctoral research fellow at University of Leeds, UK.  His current research interests include energy optimization of ICT networks, network coding and energy efficient routing protocols in optical networks. 
\end{IEEEbiographynophoto}

\begin{IEEEbiographynophoto}{Dr. Taisir Elgorashi}
received the B.S. degree (first-class Honors) in electrical and electronic engineering from the University of Khartoum, Khartoum, Sudan, in 2004, the M.Sc. degree (with distinction) in photonic and communication systems from the University of Wales, Swansea, UK, in 2005, and the Ph.D. degree in optical networking from the University of Leeds, Leeds, UK, in 2010. She is currently a Lecturer of optical networks in the School of Electrical and Electronic Engineering, University of Leeds. Previously, she held a Postdoctoral Research post at the University of Leeds (2010–2014), where she focused on the energy efficiency of optical networks investigating the use of renew- able energy in core networks, green IP over WDM networks with data centers, energy efficient physical topology design, energy efficiency of content distribution networks, distributed cloud computing, network virtualization, and big data. In 2012, she was a BT Research Fellow, where she developed an energy efficient hybrid wireless-optical broadband access network and explored the dynamics of TV viewing behavior and program popularity. The energy efficiency techniques developed during her postdoctoral re- search contributed three out of the eight carefully chosen core network energy efficiency improvement measures recommended by the GreenTouch consortium for every operator network world- wide. Her work led to several invited talks at GreenTouch, Bell Labs, the Optical Network Design and Modeling conference, the Optical Fiber Communications Conference, the International Conference on Computer Communications, and the EU Future Internet Assembly in 2013 and collaboration with Alcatel Lucent and Huawei.
 
\end{IEEEbiographynophoto}


\begin{IEEEbiographynophoto}{Prof. Jaafar Elmirghani}
is the Director of the Institute of Communication and Power Networks within the School of Electronic and Electrical Engineering, University of Leeds, UK. He joined Leeds in 2007 and prior to that (2000–2007) as chair in optical communications at the University of Wales Swansea he founded, developed and directed the Institute of Advanced Telecommunications and the Technium Digital (TD), a technology incubator/spin-off hub. He has provided outstanding leadership in a number of large research projects at the IAT and TD. He received the Ph.D. in the synchronization of optical systems and optical receiver design from the University of Huddersfield UK in 1994 and the DSc in Communication Systems and Networks from University of Leeds, UK, in 2014. He has co-authored Photonic switching Technology: Systems and Networks, (Wiley) and has published over 500 papers. He has research interests in optical systems and networks. Prof. Elmirghani is Fellow of the IET, Fellow of the Institute of Physics and Senior Member of IEEE. He was Chairman of IEEE Comsoc Transmission Access and Optical Systems technical committee and was Chairman of IEEE Comsoc Signal Processing and Communications Electronics technical committee, and an editor of IEEE Communications Magazine. He was founding Chair of the Advanced Signal Processing for Communication Symposium which started at IEEE GLOBECOM’99 and has continued since at every ICC and GLOBECOM. Prof. Elmirghani was also founding Chair of the first IEEE ICC/GLOBECOM optical symposium at GLOBECOM’00, the Future Photonic Network Technologies, Architectures and Protocols Symposium. He chaired this Symposium, which continues to date under different names. He was the founding chair of the first Green Track at ICC/GLOBECOM at GLOBECOM 2011, and is Chair of the IEEE Sustainable ICT Initiative within the IEEE Technical Activities Board (TAB) Future Directions Committee (FDC) and within the IEEE Communications Society, a pan IEEE Societies Initiative responsible for Green and Sustainable ICT activities across IEEE, 2012-present. He is and has been on the technical program committee of 38 IEEE ICC/GLOBECOM conferences between 1995 and 2019 including 18 times as Symposium Chair. He received the IEEE Communications Society Hal Sobol award, the IEEE Comsoc Chapter Achievement award for excellence in chapter activities (both in 2005), the University of Wales Swansea Outstanding Research Achievement Award, 2006, the IEEE Communications Society Signal Processing and Communication Electronics outstanding service award, 2009, a best paper award at IEEE ICC’2013, the IEEE Comsoc Transmission Access and Optical Systems outstanding Service award 2015 in recognition of “Leadership and Contributions to the Area of Green Communications”, received the GreenTouch 1000x award in 2015 for “pioneering research contributions to the field of energy efficiency in telecommunications", the 2016 IET Optoelectronics Premium Award and shared with 6 GreenTouch innovators the 2016 Edison Award in the “Collective Disruption” Category for their work on the GreenMeter, an international competition, clear evidence of his seminal contributions to Green Communications which have a lasting impact on the environment (green) and society. He is currently an editor of: IET Optoelectronics, Journal of Optical Communications, IEEE Communications Surveys and Tutorials and IEEE Journal on Selected Areas in Communications series on Green Communications and Networking. He was Co-Chair of the GreenTouch Wired, Core and Access Networks Working Group, an adviser to the Commonwealth Scholarship Commission, member of the Royal Society International Joint Projects Panel and member of the Engineering and Physical Sciences Research Council (EPSRC) College. He was Principal Investigator (PI) of the £6m EPSRC INTelligent Energy awaRe NETworks (INTERNET) Programme Grant, 2010-2016 and is currently PI of the £6.6m EPSRC Terabit Bidirectional Multi-user Optical Wireless System (TOWS) for 6G LiFi Programme Grant, 2019-2024. He has been awarded in excess of £30 million in grants to date from EPSRC, the EU and industry and has held prestigious fellowships funded by the Royal Society and by BT. He was an IEEE Comsoc Distinguished Lecturer 2013-2016.
 
\end{IEEEbiographynophoto}




%

\end{document}